\newcommand {\matr}[1] {\ushortd{\mathbf #1}}

\newcommand{\Om}{\Omega}			% Shift in frequenza
\newcommand{\q}{\vec{q}}                              %q-vector
\newcommand{\ex}{\vec{e}_x}   %Sistema di coordinate lab
\newcommand{\ey}{\vec{e}_y}

\newcommand{\etre}{\vec{e}_3}
\newcommand{\DD}{{\mathcal D}}                           % Phase mismatch general
\newcommand{\D}{{D}}                                                % Phase mismatch  plane wave pump 
\newcommand{\Dcoll}{{\mathcal D}_{\mathrm 0}}  
\newcommand{\Q}{\vec{Q}}                                         %  Vettore transverso pompa, generico  plane wave pump 
\newcommand{\Qone} { \vec{Q}_1}
\newcommand{\Qtwo} { \vec{Q}_2}
\newcommand{\gbar} {\bar{g}} % guadagno totale 
\newcommand{\Dbar} {{D}} % phase-mismatch dei modi shared  
\newcommand{\As} {\hat{A}_s}

\newcommand{\azero} {\hat{a}_{0s}}
\newcommand{\auno} {\hat{b}_{i}}
\newcommand{\adue} {\hat{c}_{i}}
\newcommand{\apiu} {\hat{d}_{i+}}
\newcommand{\ameno} {\hat{d}_{i-}}
\newcommand{\azeroi} {\hat{a}_{0i}}
\newcommand{\aunos} {\hat{b}_{s}}
\newcommand{\adues} {\hat{c}_{s}}
\newcommand{\abar} {\bar{\alpha}_{p}}
\newcommand{\phip} {\bar \phi_{p}}
\newcommand{\phim}{\phi_{-}}

\newcommand{\bs} {\hat{b}_s}
\newcommand{\bi} {\hat{b}_{i}}
\newcommand{\cs} {\hat{c}_s}
\newcommand{\ci} {\hat{c}_i}

\newcommand{\sigmas} {\hat{\sigma}_s}
\newcommand{\sigmai} {\hat{\sigma}_i}
\newcommand{\deltas} {\hat{\delta}_s}
\newcommand{\deltai}  {\hat{\delta}_i}
\newcommand{\sigmak} {\hat{\sigma}}
\newcommand{\deltak} {\hat{\delta}}

\newcommand {\chidue } { \chi^{(2)} }
\newcommand{\w}{\vec{w}}			% 3-D q,omega

\newcommand{\nn}{\nonumber}
\newcommand{\bsub}{\begin{subequations}}
\newcommand{\esub}{\end{subequations}}
\newcommand{\beq}{\begin{equation}}
\newcommand{\eeq}{\end{equation}}
\newcommand{\beqa}{\begin{eqnarray}}
\newcommand{\eeqa}{\end{eqnarray}}
\newcommand{\beql}{\begin{subequations}\begin{eqnarray}}
\newcommand{\eeql}{\end{eqnarray}\end{subequations}}
\documentclass[aps,pra,amsmath, nofootinbib,showpacs,longbibliography]{revtex4-1}
\usepackage{amsmath} 
\usepackage{empheq}
\usepackage{amssymb}
\usepackage{graphicx}
\usepackage{longtable}
\usepackage{float}   % per posizionare meglio le figure con [H]
\usepackage{verbatim}  %multi-line comment \begin{comment}....\end{comment}
\usepackage{braket}
\usepackage{xcolor}
\DeclareMathOperator{\tg}{tg}
\usepackage{ushort}
\begin{document}
\title{Engineering multipartite entanglement in doubly pumped  parametric down-conversion processes}
\author{ Alessandra~Gatti$^{1,2}$, Enrico~Brambilla$^2$ and Ottavia Jedrkiewicz$^{3,2}$}
\affiliation{$^1$ Istituto di Fotonica e Nanotecnologie (IFN-CNR), Piazza Leonardo  Da Vinci 32, Milano, Italy;  
$^2$ Dipartimento di Scienza e Alta Tecnologia dell' Universit\`a dell'Insubria, Via Valleggio 11,  Como, Italy;
$3$ Istituto di Fotonica e Nanotecnologie (IFN-CNR), Via Valleggio 11, Como, Italy }
\email{Alessandra.Gatti@ifn.cnr.it}
\begin{abstract}
We investigate the quantum  state generated by optical  
 parametric down-conversion  in a $\chi^{(2) } $ medium driven by two noncollinear light modes. 
The analysis shows the emergence of multipartite, namely 3- or 4-partite, entangled states in a subset of the spatio-temporal modes generated by the process. These appear as   bright  spots against the background fluorescence, providing an interesting analogy with the phenomenology recently observed in  two-dimensional nonlinear photonic crystals. 
We study two realistic setups: i) Non-critical  phase-matching in a periodically poled Lithium Tantalate slab, characterized by a 3-mode entangled state.   ii) A type I setup in a Beta-Barium Borate crystal,  where   the spatial walk-off between the two pumps can be exploited to make a transition to a quadripartite entangled state. In both cases we show that the properties of the state can be controlled by modulating the relative intensity of two pump waves, making the device a versatile tool for quantum state engineering. 
\end{abstract}
%\centerline{Version \today}
\maketitle
\nopagebreak
\section*{Introduction}
Multipartite entanglement, where quantum entanglement is shared by more than two physical systems, is a key resource,  both from  fundamental  \cite{Armstrong2015} and  applicative viewpoints. In optics, an efficient tool able to prepare  and engineer  multiparty entangled states of light would  be  an asset for several quantum technologies: among them,  measurement-based  quantum computation \cite{Raussendorf2001, Menicucci2006}, which require to generate  multipartite entangled cluster states 
 \cite{Briegel2001, Zhang2006} in a controlled and re-configurable way,  and quantum metrological schemes of distributed quantum sensing \cite{Shapiro2018}. Nevertheless,  the most efficient sources of  quantum optical states are nonlinear processes, as   four-wave mixing and parametric down-conversion (PDC), that  generate  photons in pairs, which naturally leads to bipartite Einstein-Podolsky-Rosen  (EPR) entanglement and to  squeezed states. Then,  in the continuous variable regime, a  well-established  scheme for producing multipartite  entanglement requires an external manipulation of such squeezed states  by mixing them in a  network of passive optical elements  (see e.g. \cite{VanLoock2007, Furusawa2008, Furusawa2013}). External manipulations of  the squeezed  or EPR states generated by nonlinear optical process are also necessary for other fundamental tasks of quantum information,  e.g.  in order to introduce non-Gaussianity and to enable entanglement distillation, as in protocols of photon-subtraction where a small fraction of the light is redirected towards a photon counting detector, and the remaining state is conditioned upon detection of photons
(see e.g.\cite{Navarrete-Benlloch2012,Takahashi2010}).

In this work we follow a different approach, aiming at engineering    the nonlinear  process  which is source of squeezing itself  by  acting on the spatial degrees of freedom of the  pump beam driving the process. Ideally,  the goal is    directly producing  the desired state and/or    implementing  some  operations of interest for quantum technologies.  In a sense, we propose to invert the order of the above mentioned steps, by transferring  the linear manipulations from the squeezed modes generated by the process to the spatial modes of the  classical laser pump beam.  This approach has a number of advantages: first, avoiding as much as possible manipulations of the fragile quantum states, operating instead on the more robust classical pump; second, the possibility of engineering   the state by modulating the properties of the pump; finally, as we shall see, the fact that multipartite entanglement is produced among  spatial modes of the same beam, which are already separated.  In this proposal we  consider using two  pump beams slightly tilted in the transverse direction to drive parametric down-conversion in a $\chidue$ medium. 

The idea is not completely new: an ideal scheme was explored in  \cite{Daems2010}, were a tripartite entanglement was theoretically predicted. The use of a spatially structured pump  with a $TEM_{01}$ modal profile, to produce peculiar spatial correlation between twin photons was also explored in \cite{Menzel2013}. More recently, a scheme for engineering the quantum and classical properties of parametric generation by dual pumping a 2D nonlinear photonic crystal  was proposed by some of us \cite{Brambilla2019,Gatti2020}. Several four-wave mixing  schemes, exploiting the $\chi^{(3)}$ nonlinearity, with dual spatial pump modes  have been recently studied and experimentally realized \cite{Wang2017,Liu2019,Zhang2020}. 

The problem with down-conversion, intrinsically more efficient than four-wave mixing, is that  both the phase matching and the effective nonlinearity depend  on the direction of propagation of each pump beam, making the scheme more complex.  Therefore, a large part of this work will be devoted to the characterization of the simultaneous phase-matching of the concurrent nonlinear processes driven by the two pumps. Two concrete setups will be explored: 
%suitable for generating multipartite entanglement by means of a structured pump. 

The first  scheme considers a type 0 process in a periodically poled Lithium Tantalate (PPLT) slab,  where the two pumps are  tilted in the plane perpendicular to the optical axis of the crystal, so that  neither the phase matching nor the nonlinearity depend on the direction of propagation of the modes. We show that this is the ideal framework to realize the proposal in \cite{Daems2010}, and that a  3-mode entanglement  is  realized in specific subsets of spatio-temporal modes, which appear as bright hot-spots against the less intense background due to standard 2-mode fluorescence.
%%%%%%%%%%% FIGURA Equivalence 3-mode -BS 
\begin{figure}[ht]
\includegraphics[scale=0.55]{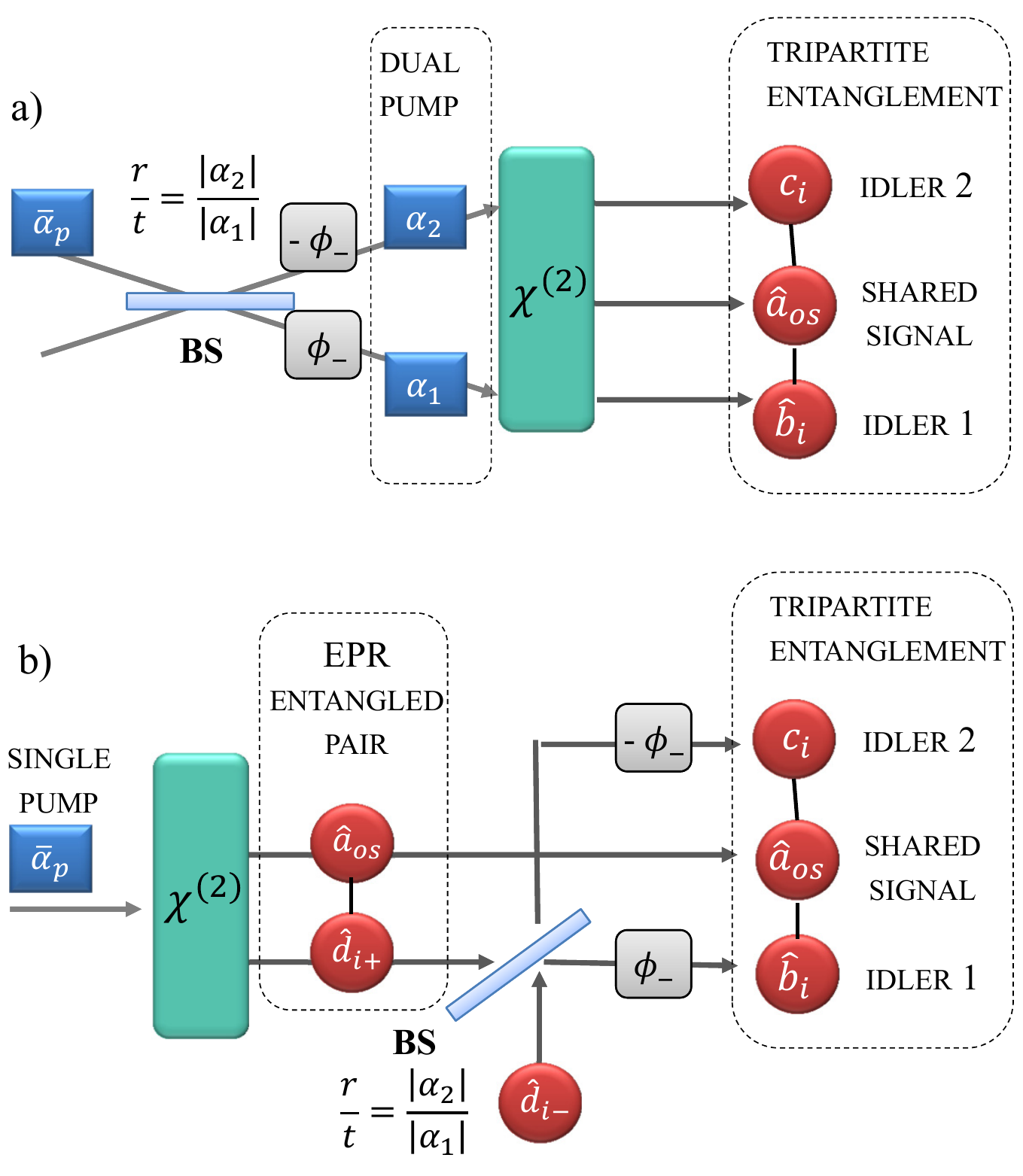}
\caption{a) Schematic of the dual pump source, where two non-collinear modes of amplitudes  $\alpha_1$ and $ \alpha_2$ pump  a $\chidue$ medium, and generate a tripartite entangled state in  specific sets of spatio-temporal modes  (Sec.\ref{sec_shared}). 
The two pumps can be seen as  deriving from a  single pump  $\abar= \sqrt{|\alpha_1|^2 + |\alpha_2|^2} $.   We will  show that    the source is formally equivalent to  the    scheme b),  in which  the medium is pumped by a  single beam of amplitude $\abar$,  and one of the two parties of the   EPR state thereby generated  is mixed with an arbitrary  input mode $\ameno $  on a beam-splitter with  reflection and trasmission coefficients  $\frac{r}{t} =  \frac{|\alpha_2|}{|\alpha_1} $.  $\phim$ are local phase shift by half of the pump  phase difference. 
%The description holds for each triplet of shared-coupled modes (see text) . 
}
\label{fig_equiv0}
\end{figure}
%%%%%%%%%%%%%%%%%%%%%%%
 With respect to \cite{Daems2010},  we analyse  the more general case  of arbitrary pump  amplitudes, and,   as schematically shown by Fig.\ref{fig_equiv0},  we  find  that  the tripartite entangled state thereby realized  is formally equivalent to  dividing  one of the parties of a bipartite EPR state 
on a beam-splitter  whose  reflection and transmission coefficients are in the same ratio as  two pump intensities.  This result may  be relevant for photon-subtraction protocols,    because   it shows that  the doubly-pumped scheme  implements  an arbitrary beam-splitter,   without the need of external alignments potentially detrimental for the quantum state. 
\par
\begin{figure}[ht]
\includegraphics[scale=0.48]{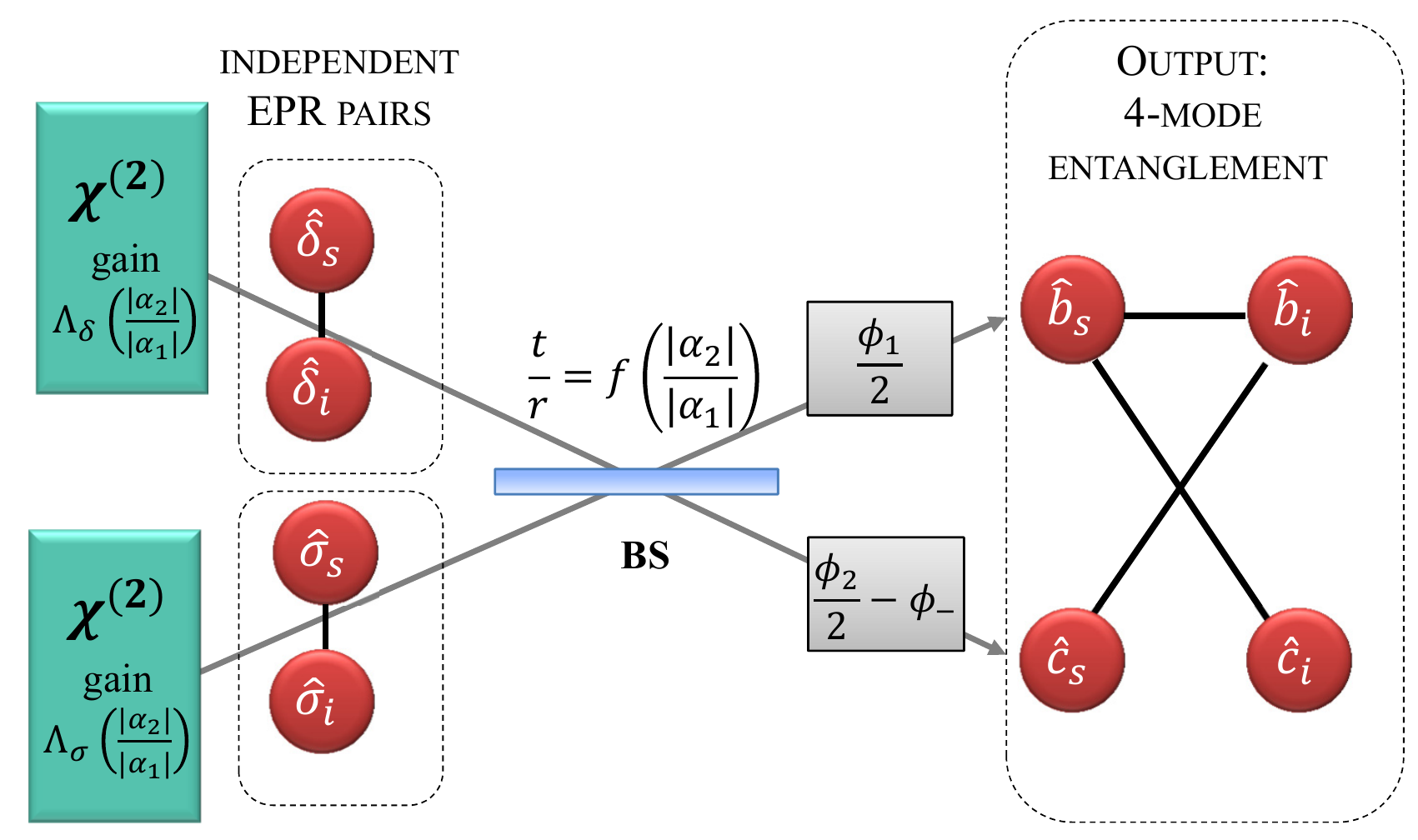}
\caption{Decomposition of the  4-mode entangled state generated by a doubly pumped BBO  in the {\em resonance} conditions demonstrated in Sec.\ref{sec_BBO}. This is equivalent to a pair of independent  EPR states mixed on a beam-splitter BS and followed by phase rotations (gray boxes). The squeeze parameters $\Lambda_\sigma$, $\Lambda_\delta$ of the EPR states,    and the  BS transmission and reflection coefficients are controlled by the intensity ratio of the two pump modes  (see Eqs.\eqref{Lambda} and \eqref{Upar}).
The description holds for specific quadruplets of spatio-temporal modes which  will be characterized in Sec. \ref{sec_transition}.
}
\label{fig_unfolding1}
\end{figure}

The second scheme considers a type I process  in a standard Beta-Barium-Borate (BBO)  crystal, where the pumps are tilted in a direction that is not perpendicular to the optical axis. The analysis here is strictly connected to a parallel experimental work \cite{Jedr2020}.
In  this configuration, we show that  the strong birefringence of the BBO crystal, responsible for  spatial walk-off effects,  can  be exploited to identify peculiar directions of propagation of the two pumps inside the crystal, such that two triplets of hot-spots, originally uncoupled,  merge into   quadruplets of entangled modes,  with  a sudden enhancement of the  intensity of hot-spots \cite{Jedr2020}.  These {\em resonances}, as we shall call them,  will be interpreted in terms of a  superposition of the mean flux of  the pump energy   (the Poynting vector of the career wave)  with the direction of propagation of  one of the pump modes. 
From a quantum viewpoint,  we will show that the quadripartite entangled state thereby  generated 
  can be formally described as the interference of  {\em a pair} of  independent EPR  states,  as schematically described by Fig.\ref{fig_unfolding1} . 
%%%%%%%%%%% FIGURA Equivalence 4-mode -BS 
%%%%%%%%%%%%%%%%%%%%%%%
 Remarkably, 
%at difference  with the NPC case,  where the properties of the state were fixed by the geometry of the crystal, we show that   in the doubly pumped BBO  scheme  
the  squeezing  and the mixing parameters  turn out to be controlled  by the relative intensity of the pumps, giving access to a potential control over the state. 

On a different perspective,  our work highlights a striking    analogy with the phenomenology recently observed in a 2-dimensional nonlinear photonic crystal (NPC) \cite{Jedr2018,Gatti2018},  including not only the  emergence of hot-spots in correspondence of triplets of entangled modes, but also the existence of the resonance that leads to a 4-mode entangled state,  which in the case of the NPC was named  {\em Golden Ratio Entanglement}.  These two apparently disparate systems have the common feature that    two  concurrent nonlinear processes coexist in the same medium, as will be discussed in Sec.\ref{sec_general}A
%Somehow surprisingly, this transition  turns out  analogous to the transition to the {\em Golden Ratio Entanglement}  recently predicted  in a nonlinear photonic crystal (NPC)\cite{Gatti2018,Jedr2020}. As  for  the NPC source,  

The paper is organized as follows: Section \ref{sec_general} introduces the general  theoretical framework and discusses the analogy between  the doubly pumped scheme and  parametric generation in nonlinear photonic crystals. Sec.\ref{sec_PPLT}  analyses the PPLT case and the tripartite entanglement associated with it, with a blend of  analytical calculations, performed in the parametric limit, and numerical simulations. 
Sec.\ref{sec_BBO} analyses the BBO case, the transition to resonance and the 4-mode entanglement. Numerical and experimental data for this part are presented in the related work \cite{Jedr2020}.
%%%%%%%%%%%%%%%%%%%%%%%%%%%%%% 
\section{General framework}
\label{sec_general}
This section introduces the general theoretical framework, formulated in terms of 
3D+1 propagation equations inside the nonlinear $\chi^{(2)}$ material for the quantum field operators associated with the interacting light fields. 

Our work focuses on a  degenerate  type 0 or type I  process, in which  the  down-converted light  is described by a single field envelope centered around half of the pump frequency. Thus, we
 consider the two slowly varying  field operators associated with the high-frequency  pump and the low-frequency  down-converted signal, which in the  Fourier domain read: 
$\hat A_j (\q,\Omega,z) = \int \frac{d^2 \vec r}{2\pi}  \int \frac{dt}{\sqrt{2\pi} } e^{ i(\omega_j + \Omega) t} e^{-i [k_{jz} (\q, \Om)z + \q \cdot \vec r ]} 
\hat E_j (\vec r, z,t)  $, ($j=p,s$) (see \cite{ Gatti2003,Brambilla2012} for details),  where:  $z$ is the mean direction of propagation of the fields, assumed to be paraxial waves; 
 %\begin{itemize} \setlength\itemsep{0em}  \item  [-]   \item [-] \end{itemize}
$\Omega_j$ is  the frequency shift from  the carriers $\omega_p$  and  $   \omega_s= {\omega_p \over 2}$; 
$\q=q_x \ex + q_y \ey $  is   the transverse component of the wave-vector; 
 $k_{jz} (\q,  \Om) = \sqrt{k_j^2 ( \q, \Om)- q^2}$  is its z-component,  where $k_j (\q,\Om)  = n_j (\q, \omega_j + \Omega) \frac{ \omega_j+ \Omega}{c}$ is the  wave-number, $n_j (\q,\omega)$ being the index of refraction of the j-th wave. For the extraordinary wave,  the index depends both  on the frequency and on the direction of propagation, implicitly identified by the transverse wave-vector component $\q$. 
  Finally,
 $\hat E_j (\vec r,z, t)$ is the full  field operator in the direct space, such that  $\hat E^\dagger _j \hat E_j$ has the dimensions of a  photon number per unit area and unit time.   
By using the shorthand notation $\w \equiv (\q, \Omega) \in \mathbb R^3$, the coupled propagation equations  have  the  form:
\bsub
\begin{align}
\frac{\partial}{\partial z}   \hat{A}_s  (\w_s, z )   &=   \int 
 \frac{d^3 \w_p }  {(2\pi)^{\frac{3}{2}} }  \chi (\w_p;\w_s)\hat{A}_p(\w_p,z) 
  \hat{A}_s^\dagger(\w_p -\w_s , z)  e^{-i \DD(\w_s; \w_p) z }  
%&\left[   \hat{b}^\dagger(\w_p -\w_s -\Gx, z)  e^{-i \DD_1 (\w_s, \w_p -\w_s-\Gx) z } \right. \nn \\ + & \left.   \hat{b}^\dagger(\w_p -\w_s +\Gx, z)  e^{-i \DD_2 (\w_s, \w_p -\w_s +\Gx) z }  \right]
\label{NLs}\\
\frac{\partial}{\partial z}    \hat{A}_p  (\w_p, z )   &= - \frac{1}{2} \int 
 \frac{d^3 \w_s }  {(2\pi)^{\frac{3}{2}} }    \chi (\w_p;\w_s) \hat{A}_s (\w_s,z) 
   \hat{A}_s  (  \w_p  -\w_s, z)  e^{i \DD(\w_s; \w_p ) z } 
% \right. \nn \\ + & \left.  
\label{NLp}
\end{align}
\label{NL}
\esub
The two equations describe all the possible down- and up-conversion processes between a pump photon in mode $\w_p$  and a  pair of signal and idler  photons  in modes $\w_s$ and $\w_i= \w_p -\w_s$,  satisfying the energy and transverse momentum conservation (for simplicity,  we assumed that the crystal is infinite in the transverse directions). The conservation of longitudinal momentum is less stringent because of the finite longitudinal size of the medium,  and is expressed by the phase-mismatch function 
\beq
\DD (\w_s; \w_p) = k_{sz} (\w_s) + k_{sz} (\w_p - \w_s) -k_{pz} (\w_p) + G_z
\label{DD}
\eeq
where we allow for the possibility of a longitudinal 1D poling of the material, such that  the reciprocal vector of the nonlinear grating $G_z = {2 \pi \over \Lambda_{pol}}$ contributes to the momentum balance.    For more generality,  we also leave the possibility for the effective nonlinearity  to  depend on the direction of propagation of the three waves, through $ \chi (\w_p, \w_s)  \propto d_{eff} ( \w_p, \w_s) \sqrt{\frac{ \hbar \omega_p  \omega_s^2   }{ 8\epsilon_0 c^3 n_e (\omega_p) n_e^2(\omega_s)  }}  $. In standard configurations, where the pump is a weakly focused Gaussian beam propagating around a single direction,  this dependence can be neglected . When the pump  transverse profile is structured, in particular when it is formed  by  several waves propagating at different angles, the effective nonlinearity can significantly differ in each direction.
\par
The nonlinear  equations \eqref{NL}  have been numerically simulated, by means of 
 fully 3D +1 simulations (see \cite{Gatti1997} and methods of \cite{Jedr2018}), which take  into account a broad frequency bandwidth, typically  on the order of 200-400 nm. 
The pump modes are modelled by two 
Gaussian pump pulses, of duration $ \simeq 1$ ps and transverse waist $\simeq 400 \mu$m,   which propagate  close to  the $z$ axis tilted one with respect to the other by  few degrees. 
\subsection{Multiple pump waves, analogy with Nonlinear Photonic Crystals}  
Although numerical simulations can fully  account for pump depletion effects, in the remaining of this work we shall largely exploit  the  undepleted pump limit. Thus we focus on  Eq.\eqref{NLs} only,   with 
the  pump field operator   $ \hat A_p (\w_p, z )$     replaced by  the  classical envelope  $ {\cal A}_p ( \w_p) $ describing  the  profile of the  injected pump. 
In particular, we  consider the injection of multiple plane-wave modes, propagating at slightly different directions around the $z$-axis, i.e. 
\beq 
{\cal A}_p (\q,\Omega) = (2 \pi)^{\frac{3}{2}} \delta (\Omega= 0)   \sum_m \alpha_m \delta (\q - \Q_m) 
\eeq
Neglecting for simplicity  in Eq \eqref{NLs}  the dependence of the effective nonlinearity on the propagation direction, we can  then build a  straightforward analogy with the process of parametric generation in 2D nonlinear photonic crystals \cite{Berger1998, Broderick2000}. In such materials,  the nonlinear response of the medium is artificially modulated, typically via ferroelectric poling,  according to a 2D periodic pattern, the pattern  lying in a plane $(x,z)$ perpendicular to the optical axis, including thus one transverse direction. 
%the pattern lying  in the $(x,z)$ plane perpendicular to the optical axis.  
The transverse  modulation of the nonlinear  response can be often reduced  to $ \chi (q_x) \to 
 \sum_m  \chi_{m}  \delta( q_x - \vec G_{m}) $, where $\vec G_{m}$ are the transverse components of the reciprocal vectors of the nonlinear lattice participating to quasi phase-matching \cite{Arie2007}. For example, for a hexagonally poled crystal $ \vec G_m= \pm G \ex$ are the transverse components of the two fundamental reciprocal  lattice vectors (see \cite{Jedr2018,Gatti2018,Gatti2020}).   According to Eq.\eqref{NLs},   it  is  therefore equivalent  to inject a  single plane-wave pump  into a photonic crystal  equipped with  several non-collinear  reciprocal lattice vectors, or  to inject   several non-collinear pumps into a  standard crystal (or  into a 1D poled crystal). In a less formal way,  the down-conversion process from an undepleted pump beam is ruled by the product of the medium nonlinear response and the pump profile: thus it is equivalent to structure the transverse profile of either one or the other.  In the following of this work we shall indeed show that the behaviour of a nonlinear photonic crystal can be fully mimicked by injecting two non-collinear pumps, with the additional benefit that the dual pump scheme enables a reconfigurable control over the properties of the process, by modulating the pump amplitudes. 
%%%%%%%%%%%%%%%%%% Section II 
\section{Type 0 process: tripartite entanglement} 
\label{sec_PPLT}
%%%%%%%%%%% FIGURA SETUP fig_setup
\begin{figure}[ht]
\includegraphics[scale=0.6]{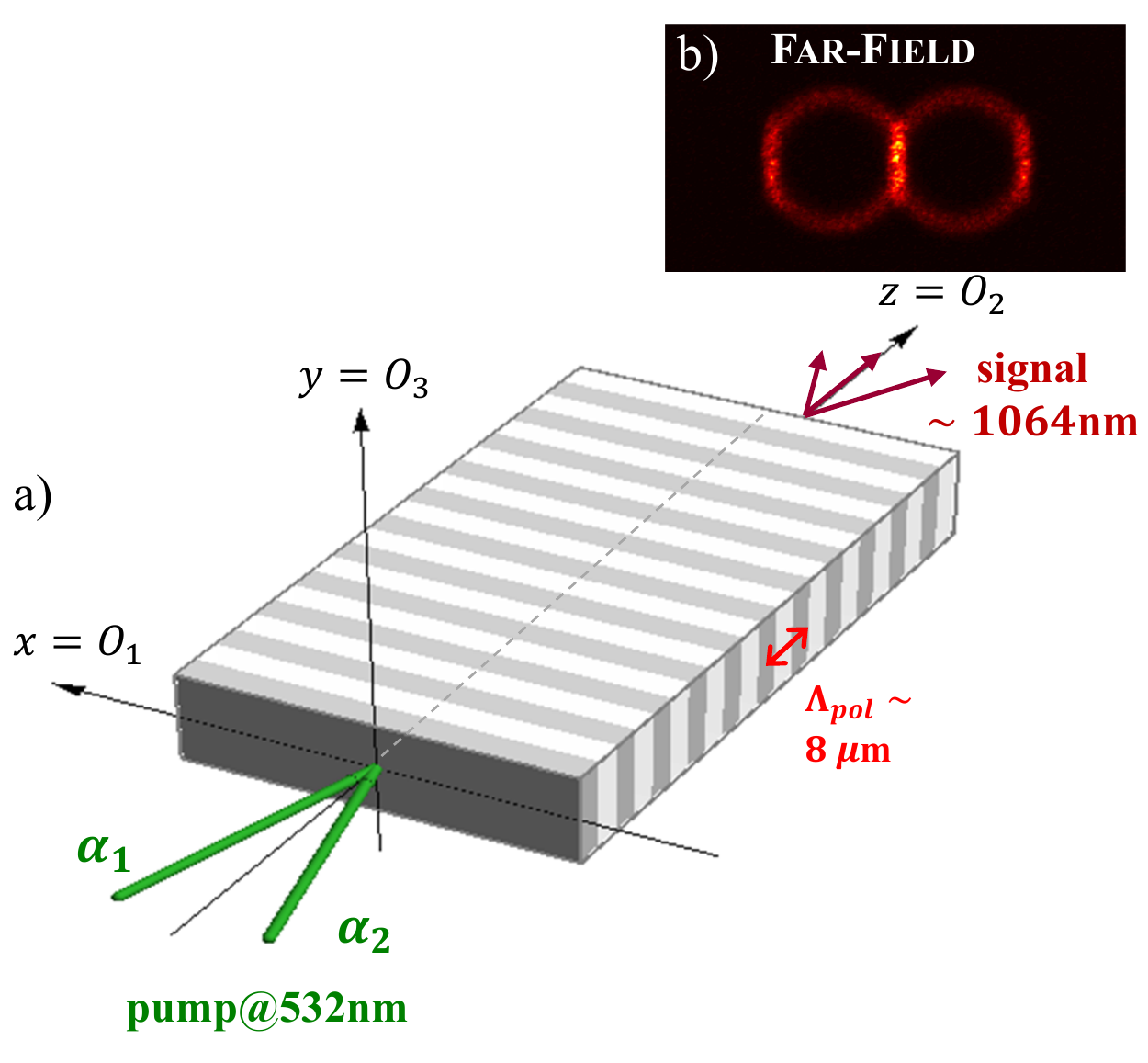}
\caption{ a)Geometry of the scheme for the e$ \to $e,e  process in a periodically poled LiTa0$_3$ slab. $O_1,O_2,O_3$  are the crystal principal axes.   All the fields are polarized along the optical $O_3$ axis, and propagate at small angles with  the $O_2$ axis. The  pumps are slightly tilted along  $x=O_1$. b) Far-field distribution of down-converted light from numerical simulations (see text). The plot shows a $20$nm bandwidth around $1064$nm.
}
\label{fig_setup0}
\end{figure}
%%%%%%%%%%%%%%%%%%%%%%%
This section  studies the simplest configuration: a type O process where all the waves are extraordinarily polarized, pumped by two beams  that propagate noncollinearly in the plane perpendicular to the optical axis. 

For definiteness, we consider a periodically poled  LiTa0$_3$  slab\footnote{The analysis can be straightforwardly extended to  PPLN, we choose LiTaO$_3$ as an active material because of its very small birefringence}, with a poling period $\Lambda_{pol} \simeq 7.9 \mu$m,  suitable to phase-match the type O interaction   $\lambda_p = 532$nm $ \to \lambda_s = \lambda_i  = 1064$nm at a temperature of $T \approx 75^\circ$. 
 Fig.\ref{fig_setup0} shows the geometry of the scheme: $O_3 \equiv y $ is  the optical axis of the crystal;   the crystalline $O_2 \equiv z $ axis represents the mean propagation direction  of all  waves;   the two injected pump waves are slightly  tilted in the $O_1\equiv x $ direction, and thus  propagate in the plane perpendicular to the  optical axis. In these conditions, also known as {\em non-critical phase matching}, their wave numbers do not depend on the tilt angle, and,  assuming a  paraxial propagation, also the nonlinear coefficient does not depend to a good approximation  on their propagation directions. 

\begin{figure*}
\includegraphics[scale=0.65]{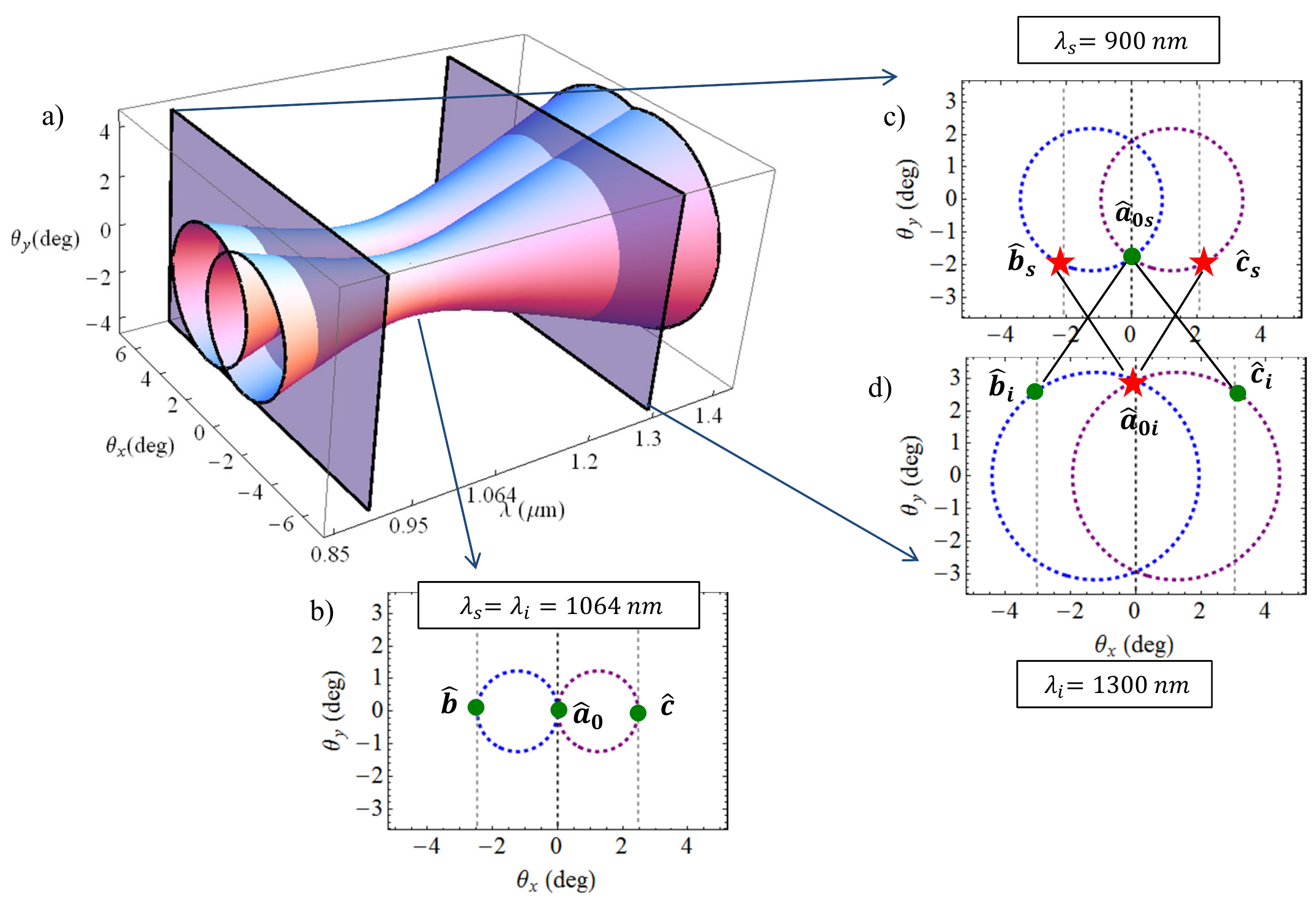}
\caption{PPLT doubly pumped at $532$nm, by two pumps tilted at $\theta_{p1,p2} = \mp 1.2^\circ$ (internal angle).  (a) Surfaces in the $(\lambda,\theta_x,\theta_y)$ space defining phase-matching for each pump mode  [see Eq.\eqref{Sigma0}],  calculated with the Sellmeier formulas in \cite{Dolev2009}, for T=75$^0$ and $\Lambda_{pol}= 7.79 \mu$m. (b) Section at   $\lambda_s=\lambda_i=1064$nm, showing the triplet of entangled modes. (c) and (d): Sections at two  conjugate wavelengths $\lambda_s=900$nm, $\lambda_i=1300$nm, showing two independent triplets of modes (stars and dots).
}
\label{fig_PM0}
\end{figure*}

We approximate the pumps  as classical  plane-waves of complex amplitudes $\alpha_1$ and $\alpha_2$,  characterized by  transverse wave vectors $\Qone = Q_1\ex$ , $\Qtwo = Q_2 \ex $ , where $|Q_m | \ll {2\pi\over \lambda_p}$. 
% such that the pump transverse envelope has the form \beq { \cal A}_p (x) = \alpha_1 e^{i Q_1 x} + \alpha_2 e^{i Q_2 x} \label{Apx} \eeq
By substituting into Eq. \eqref{NLs}, we get: 
\begin{align}
\frac{\partial}{\partial z} \hat A_s (\w, z) &= \chi \left[ \, \alpha_1 \hat A_s^\dagger ( \Qone-\w, z) e^{- i \DD  (\w; \Qone) z}  \right.  \nn  \\
&  + \left.  \, \alpha_2 \hat A_s^\dagger (\Qtwo-\w ,z)  e^{- i \DD  (\w; \Qtwo) z} \right] ,
\label{prop}
\end{align} 
where $\chi \simeq \chi (\w_{p1};\w) =\chi (\w_{p2};\w)$ is the common value of the nonlinear coefficient. The  r.h.s of Eq. \eqref{prop}  shows the contribution of the two  processes originating from each  pump.  For the large majority of modes,  only one of the two processes is phase-matched, giving rise to two  noncollinear  branches of down-converted modes (examples are shown in Fig. \ref{fig_PM0}), corresponding to the standard conical emission around each pump taken separately.  In the  Fourier space $(\q, \Omega)$ photon pairs down-converted from each pump   populate  surfaces of equation
\beq
\begin{aligned}
&\Sigma_1 :   \; \DD  (\w; \Qone)=0 ,  &\text{ pump 1} \\
&\Sigma_2 :   \; \DD  (\w; \Qtwo)=0  &\text{ pump 2}
\label{Sigma0}
\end{aligned} 
\eeq
%%%%%%%%%%% FIGURA QPM and shared-modes 
%%%%%%%%%%%%%%%%%%%%%%%
A light mode $\w$ belonging to the branch  $\Sigma_1$  ($\Sigma_2$), but not to the intersection $ \Sigma_1 \cap
\Sigma_2$,   hosts signal photons  down-converted from  pump 1 (2), whose twin idler photon  is generated in a single  coupled mode $\Qone -\w$  ($\Qtwo -\w$), giving rise to the standard two-mode coupling of PDC. 
Conversely, the modes lying at the geometrical  intersection $ \Sigma_1 \cap
\Sigma_2$ are special,  because here  phase-matching is simultaneously realized for both pumps. Therefore, a  photon  appearing in one of these {\em shared} modes has  been down-converted from either pump 1 or 2,  indistinguishably. Its twin photon appears in either one of  {\em  two coupled} modes, which  evolve according to: 
\begin{align}
\frac{\partial}{\partial z} \hat A_s (\Qone-\w, z) &= \chi \left[ \, \alpha_1 \hat A_s^\dagger ( \w, z) e^{- i \DD  (\w; \Qone) z}  \right.  \nn  \\
&  + \left.  \, \alpha_2 \hat A_s^\dagger (\Qtwo-\Qone +\w ,z)  e^{- i \DD  (\Qone -\w; \Qtwo) z} \right] ,
\label{prop1} \\
\frac{\partial}{\partial z} \hat A_s (\Qtwo-\w, z) &= \chi \left[ \, \alpha_2 \hat A_s^\dagger ( \w, z) e^{- i \DD  (\w; \Qtwo) z}  \right.  \nn  \\
&  + \left.  \, \alpha_1 \hat A_s^\dagger (\Qone-\Qtwo +\w ,z)  e^{- i \DD  (\Qtwo -\w; \Qone) z} \right] ,
\label{prop2} 
\end{align} 
where we used the fact that $\DD (\vec{Q}_m - \w; \vec Q_m) = \DD (\w; \vec Q_m)$, $ (m= 1,2)$, implicit in the definition \eqref{DD} of the phase-mismatch function.  In the present configuration, as  shown in Sec.\ref{sec_shared},   if the shared mode condition 
\beq
\DD  (\w_0; \Qone)= \DD  (\w_0; \Qtwo)  \simeq 0 
\label{shared} 
\eeq
is satisfied for a  mode $\w_0$, then the second of the two processes appearing at r.h.s of Eqs. \eqref{prop1} and \eqref{prop2} is  not phase matched, that is, 
$\DD  (\Qtwo -\w_0; \Qone) $ and $ \DD  (\Qone -\w_0; \Qtwo) $ are significantly different from zero. In other words, if the mode $\w_0$ is shared, then its two coupled modes cannot be themselves shared.   This leads to  the tripartite entangled state that will be described  in the next section.
%%%%%%%%%%%%%%%%%%%%
\subsection{Tripartite entanglement} 
\label{sec_tripartite}
Let us concentrate on a specific triplet of  modes whose coordinates $\w_0 $ (shared mode) and $\w_{b,c} = \Q_{1,2} - \w_0 $  (modes coupled to $\w_0$ via pump 1 and 2, respectively) are a  solution  of Eq.\eqref{shared}, as for example the modes shown by the dots in Fig.\ref{fig_PM0}c,d.  Indicating by $\azero := \As (\w_0)$,  $\auno: = \As (\Qone -\w_0)$, $\adue : =  \As (\Qtwo -\w_0)$ the three field operators involved,  Eqs.\eqref{prop},\eqref{prop1} and  \eqref{prop2} lead to the 3-mode  evolution: 
\bsub
\label{3prop}
\begin{align}
&\frac{d}{dz} \azero (z)  =  \chi \left[ \alpha_1 \auno^\dagger (z) + \alpha_2 \adue^\dagger (z) \right] e^{-i \D (\w_0) z} 
\label{a0}\\
&\frac{d}{dz} \auno (z)  =  \chi \left[ \alpha_1 \azero^\dagger (z) \right] e^{-i \D (\w_0) z} 
\label{a1} \\
&\frac{d}{dz} \adue (z)  =  \chi \left[ \alpha_2 \azero^\dagger (z) \right] e^{-i \D (\w_0) z} 
\label{a2} 
\end{align} 
\esub
where $\D  (\w_0) = \DD (\w_0; \Qone)= \DD (\w_0; \Qtwo)$ is the common value of the phase-mismatch.  
Eqs.\eqref{3prop} can be readily solved by means of a   linear transformation acting on the 2 side modes: 
\beq
\begin{pmatrix}
\apiu \\[0.3em]
\ameno 
\end{pmatrix}  =
% \frac{1}{ \abar }
\begin{pmatrix}
\frac{ \alpha_1^*}{\abar^*} \;  & \frac { \alpha_2^*} {\abar^*} \\[0.5em]
-\frac{ \alpha_2}{\abar}  \; & \frac{ \alpha_1}{\abar}
\end{pmatrix} 
 \begin{pmatrix}
\auno \\[0.3em]
\adue
\end{pmatrix}   
\label{trf1}
%\qquad  \abar = e^{i \frac{\phi_1 + \phi_2} {2}} \, \sqrt{ |\alpha_1|^2 + |\alpha_2|^2}. \NOTA{fine alt.} 
\eeq
%\beq
%\begin{aligned}
%\apiu &= \frac { \alpha_2^* \adue + \alpha_1^* \auno} {\abar^*}   \\
%\ameno &= \frac{\alpha_1 \adue - \alpha_2 \auno} {\abar},     
%\end{aligned} 
%\label{trf1}
%\eeq
where
\beq
\abar = e^{i \frac{\phi_1 + \phi_2} {2}} \,   \sqrt{ |\alpha_1|^2 + |\alpha_2|^2}, 
\label{abar}
\eeq
 can be seen as the complex amplitude of a single pump carrying the sum  of the  energies of the two pumps  ($\phi_1$, $\phi_2$ being  the phases of each  wave). 
As it can be immediately verified,  the new modes evolve according to: 
\bsub
\label{2prop}
\begin{align}
\frac{d}{dz} \azero (z) & =  \chi \abar \, \apiu^\dagger (z)  e^{-i \D z} 
\label{a02}\\
\frac{d}{dz} \apiu (z) & =  \chi \abar \;  \azero^\dagger (z)  e^{-i \D z} \, , 
\label{apiu} 
\end{align}
\esub 
while
%where  \beq
%\chi e^{i \phip} \sqrt{ |\alpha_1|^2 + |\alpha_2|^2}, 
%\label{Lambda} 
%\eeq 
\begin{align}
\frac{d}{dz} \ameno (z) & =  0 . \qquad \qquad \qquad \text{  }
\label{ameno} 
\end{align} 
Eqs.\eqref{2prop}  describe a standard  PDC process  involving modes $ \azero$ and $\apiu$, pumped by a  single wave  $\abar$ of energy 
$ |\abar|^2 = |\alpha_1|^2 + |\alpha_2|^2 $ and  phase  $\phip= {\phi_1 + \phi_2 \over 2}$.  As well known, the solution of Eq.\eqref{2prop}, starting from initial conditions  $ \azero^{\mathrm in}, \apiu^{\mathrm in}$ 
at the crystal entrance face,  are  Bogoliubov transformations, which for  $\D (\w_0)=0 $  (phase-matched modes)  take the  form: 
\beq
\begin{aligned}
\azero (z) &=  \cosh{(\gbar z)}  \, \azero^{\mathrm in} + e^{i \phip}  \sinh{(\gbar z)}  \, \apiu^{{\mathrm in} \dagger}  ,  \\
\apiu (z) &=  \cosh({ \gbar z)}  \, \apiu^{\mathrm in} +   e^{i \phip}  \sinh{(\gbar z)}  \, \azero^{{\mathrm in}\dagger}  .
\end{aligned} 
\label{Bogo1}
\eeq
(the case of arbitrary mismatch   can be e.g. found in the Appendix A of Ref.\cite{Gatti2018}, substituting the parameter $\gamma g_0 l_c$ appearing there with 
$ \chi | \abar |  z$).  
If instead of  the fields,   the quantum state is evolved  along the medium,  the joint state of modes $\azero, \apiu$ is  the  { EPR state} (or {\em two-mode squeezed state}, see e.g. \cite{Knight2005a}), with squeeze parameter  $\chi \abar z$.  Conversely, mode $\ameno$ does not evolve along the crystal, and its state remains  the same    it had at the crystal input (e.g.  vacuum or a coherent state): $\ameno (z)= \ameno^{\mathrm in}$.
% generated by the action of the two-mode squeeze operator
%$ \hat R_\sigma =\exp {[\zeta_\sigma (z)\sigmas^\dagger \sigmai^\dagger - \zeta^*_\sigma (z)  \sigmas \sigmai ]}$, and 
%$\hat R_\delta= \exp {[\zeta_\delta (z)  \deltas^\dagger \deltai^\dagger - \zeta^*_\delta(z)  \deltas \deltai ]}$. The
% squeeze parameters $\zeta_{\sigma, \delta}  (z)  \in \mathbb{C}$  can be calculated   in the  general case  (Appendix A of \cite{Gatti2018});  at perfect phase-matching they reduce to $\zeta_{\sigma, \delta} (z) = \Lambda_{\sigma,\delta} z \in \mathbb{R} $. 
\par 
On the other hand, by inverting  the  unitary transformation \eqref{trf1} one has $ 
\left( \begin{smallmatrix}
\auno \\
\adue
\end{smallmatrix}\right) =
\left(  \begin{smallmatrix}
\frac{\alpha_1}{ \abar   }
& - \frac{\alpha_2^*}{ \abar^*   } \\
 \frac{\alpha_2}{ \abar   } & \; \frac{\alpha_1^*}{ \abar^*   }, i.e. 
\end{smallmatrix} \right)
\left(  \begin{smallmatrix}
\abar    \\
0
\end{smallmatrix}   \right)
$, i.e. 
\beq
\begin{pmatrix}
\auno \\
\adue
\end{pmatrix}  =\begin{pmatrix}e^{i \phi_-}  & \;  0 \\
0&  e^{- i \phi_-} \end{pmatrix} \begin{pmatrix}
 \frac{|\alpha_1| }{|\abar|} \; & - \frac{|\alpha_2| }{|\abar|}  \\[0.5em]
 \frac{|\alpha_2| }{|\abar|}    \;  &  \, \frac{|\alpha_1| }{|\abar|}
\end{pmatrix} 
 \begin{pmatrix}
\apiu    \\
\ameno 
\end{pmatrix}   
\label{trf3} 
\eeq
where  $\phim =\frac{\phi_1 -\phi_2}{2}$.  It can be immediately recognized that the  transformation \eqref{trf3}  is  equivalent to the action of a lossless beam-splitter,  with transmission and reflection coefficients $t =\frac{|\alpha_1| }{|\abar|}$ and  $r=-\frac{|\alpha_2| }{|\abar|}$, respectively. 
Thus, for each triplet of  entangled modes, the  doubly pumped  PDC scheme can be considered  formally equivalent  to the sequence shown in  Fig.\ref{fig_equiv0}b, i.e. to:
\\
 i) A standard  parametric process, pumped by a single pump  of amplitude $\abar$, carrying 
 the same total energy of  the two pumps, 
generating  a  {  EPR entangled state} in modes $\azero$ and  $\apiu$.\\
 ii) A beam-splitter mixing one of the twin beams generated in step i) with an independent    input beam $ \ameno = \ameno^{\mathrm in} $; the   reflection and transmission  of the beam-splitter  are in the same ratio as the two pump amplitudes:  
$ \frac{|r|}{|t|}= \frac{ \sin \theta }{ \cos\theta }=\frac{ |\alpha_2 |} { |\alpha_1|}$;\\
 iii) Local phase shifts on the two outputs  by the half-difference of the pump phases:    $ \auno, \adue \to \auno e^{i \phim}, \,  \adue e^{-i \phim}$. 
\par
The two pumps can be in principle derived from a single pump of complex amplitude $\abar$, through the same linear transformation described by Eq.\eqref{trf3}: 
$ 
\left( \begin{smallmatrix}
\alpha_1 \\
\alpha_2 
\end{smallmatrix}\right) =\left( \begin{smallmatrix}e^{i \phi_-}  &   0 \; \;  \; \\
0\; \; \;  &  \; e^{- i \phi_-} \end{smallmatrix}\right)
\left(  \begin{smallmatrix}
\cos \theta  \; & -\sin \theta  \\
\sin \theta    \;  & \cos \theta  
\end{smallmatrix} \right)
\left(  \begin{smallmatrix}
\abar    \\
0
\end{smallmatrix}   \right)
$ (also in practice this is  a method to obtain the two pump modes\cite{Jedr2020}).
% in a Mac Zehender -like  interferometric scheme, with a rotating mirror imparting the relative tilt,
We have  shown  that the doubly pumped  source  formally implements the  same linear transformation  on one of the two parties of an EPR state,  where the other mode $\ameno$ can be in principle  externally supplied  in any arbitrary state. 
Such a splitting-mixing  can be  of relevant practical applications, in protocols of photon-subtracted  Gaussian states \cite{Navarrete-Benlloch2012,Takahashi2010}): the device gives the possibility of redirecting a portion of one party of the EPR state in a separate spatial mode  without the need of external alignments, potentially detrimental for the quantum state. The same operations are instead performed on the less fragile classical laser pump. 
\par 
This  tripartite state represents a generalization of the state studied in \cite{Daems2010} to arbitrary pump amplitudes: as that one it shows genuine tripartite entanglement, as will be discussed in a separate work \cite{Gatti2020c}. On a different perspective, it is also analogous to the state   produced by a nonlinear photonic crystal  whose quantum correlations were extensively analyzed in Ref. \cite{Gatti2018}, with the important difference that in the dual pump scheme the splitting ratio $\frac{r}{t} = \frac{|\alpha_2|}{|\alpha_1|}$ can be easily reconfigured, whereas in the NPC case it is fixed 
by the geometry of the nonlinear grating.   A further  analogy  concerns the presence of hot-spots in the parametric emission at the location of shared and coupled modes, due to the fact that their  parametric  gain   $\gbar  = \sqrt{ g_1^2 + g_2^2} $ is  larger than that of  the  surrounding two-mode fluorescence from pump 1 alone (gain   $g_1 = \chi |\alpha_1|$), or pump 2 alone  (gain $g_2= \chi |\alpha_2|$ ).  
%%%%%%%%%%% FIGURA Equivalence 3-mode -BS 
\begin{figure*}
\includegraphics[scale=0.72]{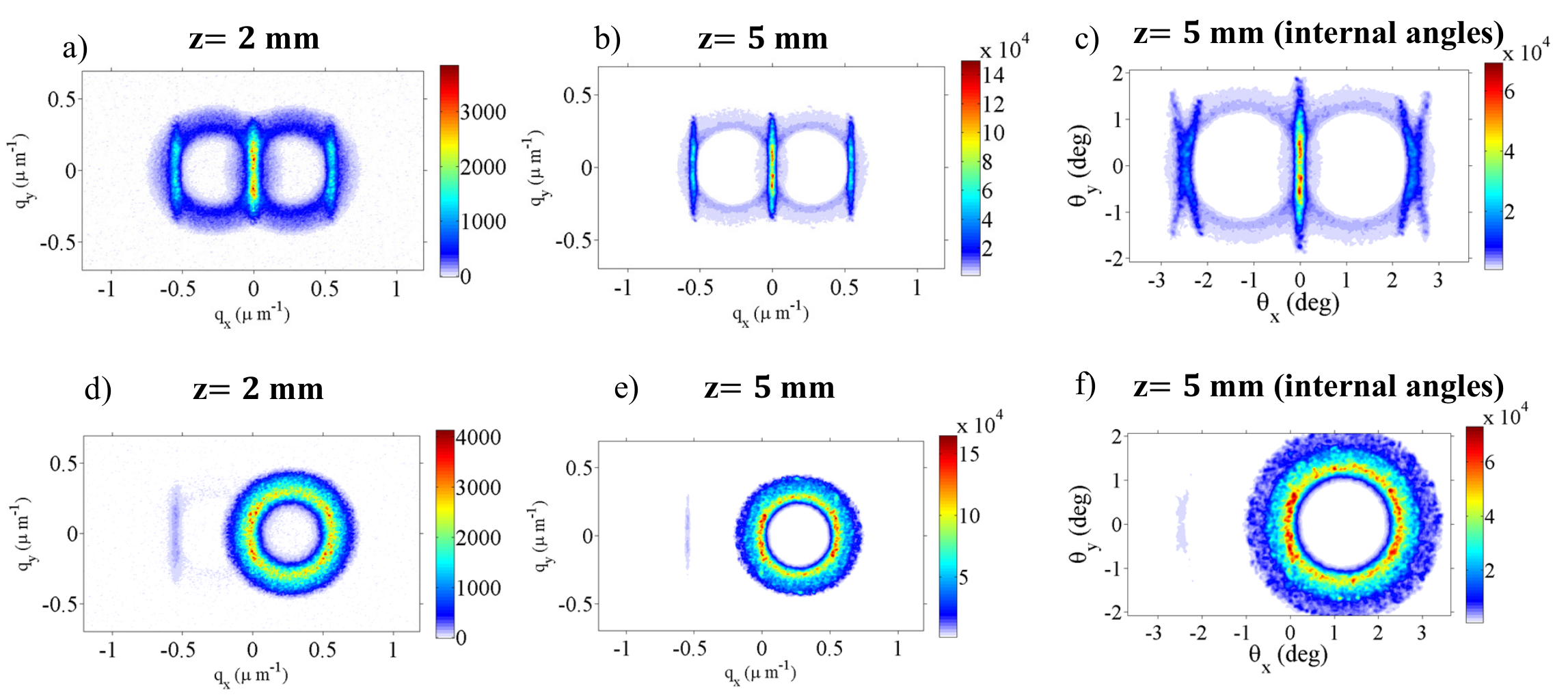}
\caption{Numerical simulations of  Eq.\eqref{NL} for a doubly pumped PPLT, showing the Fourier $(q_x,q_y)$ and angular $(\theta_x, \theta_y)$  intensity distributions of light  downconverted in the bandwidth $ 950-1210$ nm, from two  Gaussian pump pulses of peak amplitudes $\alpha_1$ and $\alpha_2$,    $1$ps duration and  $400 \, \mu$m waist, tilted at $\theta_{p1,p2} = \mp 1.2^\circ$.   In the upper row a)-c)   $\alpha_2=\alpha_1$,  with the 3 hot-spot branches becoming progressively brighter for increasing propagation length. In the lower row d) - f)  $\alpha_2=4\alpha_1$, and  the left  hot-spot branch is much weaker  than the right one.  The overall peak gain is $\gbar= \sqrt{g_1^2 + g_2^2}  =1.2 \text{ mm}^{-1}$, other parameters as in Fig.\ref{fig_PM0}.
}
\label{fig_simulazioni0}
\end{figure*}
%%%%%%%%%%%%%%%%%%%%%%%
 Therefore in the stimulated regime of PDC where the intensity grows exponentially with the gain,  these modes  appear as bright spots against a less intense background.  This is especially true when the two pumps are balanced, as shown by the simulations in the upper row of Fig.\ref{fig_simulazioni0}, performed with two Gaussian pump pulses of equal peak amplitudes. 
Notice that in these plots all the spectral components within a rather large bandwidth  are superimposed, resulting in three continuous branches of hot-spots in the source far-field.  According to the results of  the plane-wave model, their  exponential growth rate along the medium is   $\sqrt 2$  times larger than that of the background conical emission from each pump, in complete analogy with what observed in NPC sources \cite{Levenius2012,Chen2014,Jedr2018}. The case of two strongly unbalanced pumps is illustrated by the second raw of Fig.\ref{fig_simulazioni0}, where $g_2=4 g_1$:  then, the fluorescence from pump 1 is basically not visible on the scale of the plot, while the left  hot-spot branch (corresponding to modes labelled as $\hat b$ in the previous section)  is  visible, although $\sim16$ times less intense than the right branch. 
\par
A final remark concerns  the transformation \eqref{trf1} that decouples the 3-mode evolution, and its connection with the  near-field  distribution  of  modes.
The simplest case is that of symmetric pump tilts $Q_2= -Q_1$, in which the   transverse modulation of the pump occurs    along   the $x$ axis: 
 $ { \cal A}_p (x) = \alpha_1 e^{i Q_1 x} + \alpha_2 e^{-i Q_1 x} $
(the general case, in which the pump is modulated along an axis inclined at ${\theta_{p1} + \theta_{p2}} \over 2 $ is also not difficult to treat). Then, as shown in App.\ref{app_shared}, shared modes are generated  at $q_{0x}=0 $, and 
because of transverse momentum conservation, the  side modes have Fourier coordinates  $\q_{b}= Q_1  \ex $ and  $\q_{c}= -Q_1 \ex $ . They  generate a transverse field  distribution of the form: 
\beq
  \auno e^{iQ_1 x} + \adue e^{-iQ_1 x}   =
\apiu u_p (x) + \ameno u_p^{\perp} (x)  , 
\eeq
%\NOTA{ The simplest case is that of   symmetric pump tilts $Q_1 + Q_2=0 $. Then, as shown in App.\ref{app_shared}, shared modes propagate in the symmetry plane between the two pumps at $q_{0x}=0 $. The two side modes have $q_{bx} = Q_1$, $q_{cx} = Q_2$ and generate a transverse  x distribution of the form 
%$$\auno e^{iQ_1 x} + \adue e^{iQ_2 x} =  \apiu f_p (x) + \ameno f_p^{\perp} (x) , $$}
where we used Eq.\eqref{trf3}  and 
$
u_p (x)  = \frac{   \alpha_1 e^{i Q_1 x} + \alpha_2   e^{-i Q_1 x}  } {\abar}  $ can be recognized as 
 the pump  spatial mode, such that the pump envelope is ${ \mathcal A}_p (x) = \abar u_p (x) $.  
$u_p^\perp (x)  = \frac{-\alpha_2^* e^{i Q_1 x} + \alpha_1^*e^{-i Q_1 x  }}{\abar^*} $ is  the orthogonal spatial mode,  having the smallest spatial superposition to the pump $\int dx u_p^* (x)u_p^\perp (x) =0$
\footnote{Here we disregard details related to mode normalization and the finite size of the medium, which could be easily fixed by standard methods}. This makes clear the decomposition in Eq. \eqref{trf1}:    $\apiu $ is the spatial mode of the pump, and it is the only one to be parametrically amplified, while $\ameno$ is the spatial mode orthogonal to the pump and it is not affected by the parametric generation.  Notice that the result is less trivial than it might appear:  if both the pump modes were not simultaneously phase matched, then it wouldn't hold true.
%%%%%%%%%%%%%%%%%%%%%%%%%%%%%%%%%
\subsection{ Position of shared and coupled modes} 
\label{sec_shared}
%%%%%%%%%%%%%%%%%%%%%%%%%%%%%%%%%55
The tripartite entangled state studied in the previous section concerns all the triplets of shared and coupled modes that are solutions of Eq.\eqref{shared}.
Their Fourier coordinates are studied  in App.\ref{app_shared}  by using   the paraxial approximation, and are for example shown by the numerical simulations of Fig.\ref{fig_simulazioni0}. These results  can be  mapped  into
angles of propagation around the z-axis,   $  q_x \to k_s (\Omega)  \sin \theta_x \simeq 
 k_s (\Omega) \theta_x $, $  q_y \to k_s (\Omega)  \sin \theta_y \simeq 
 k_s (\Omega) \theta_y $, where   $k_s (\Omega) \simeq n_e (\omega_s + \Omega, \pi/2) \frac{\omega_s + \Omega}{c}$
\footnote{ Notice that when  propagating at nearly $\pi/2$ angle with the optical axis, the  dependence of $n_e$ on the propagation direction 
 is negligible, especially  for  LiTaO$_3$, whose birefringence  is very small \cite{Dolev2009}.}. We find that shared and coupled modes  at the frequency $\omega_s +\Omega $ are generated at angles
\begin{align}
\theta_{0x}  (\Omega)& =  { \theta_{p1} + \theta_{p2}  \over 2} \left( 1 + \frac{G_z - \Dcoll (\Omega)}{k_s (\Omega)} \right) \simeq { \theta_{p1} + \theta_{p2}  \over 2} 
\label{shared_theta}\\
\theta_{b,c x}  (\Omega) &=  { \theta_{p1} + \theta_{p2}  \over 2}   \pm  { \theta_{p1} - \theta_{p2}  \over 2}  \frac{k_p}{k_s (\Omega) }
\simeq { \theta_{p1} + \theta_{p2}  \over 2}   \pm  {( \theta_{p1} - \theta_{p2} )}  \frac{1}{1 + \Omega/\omega_s}
\label{coupled_theta}
\end{align}
where $\theta_{p1,p2}= { Q_{1,2} \over k_p}$ are the (internal) angles formed by the  two pumps with the $z$-axis, and $ \Dcoll (\Omega) = k_s (\Omega) + k_s (-\Omega) -k_p + G_z$ is  the collinear phase-mismatch parameter (i.e. the mismatch one would have if the 3 waves propagated collinearly along the z-axis). 
As  it could be  expected,  shared  modes are approximately emitted in the symmetry plane between the two pumps.
This is exactly true when $\theta_{p1} + \theta_{p2} =0 $, i.e. the career of the pump field propagates along $z$, but approximately holds also when  the tilts are not symmetric, because  $\frac{G_z} {k_s} \simeq\frac{ \lambda_s  }{n_e (\lambda_s) \Lambda_{pol}} \approx 0.06$, and   $\Dcoll \ll k_s $ \footnote{When the tilts are not symmetric, the pump career  propagates at a  slightly oblique direction with respect to the poling, and in the frame of reference aligned with the pump career, there is a  transverse contribution of the reciprocal poling vector. }.
The side modes are approximately displaced by  $\pm  ( \theta_{p1} - \theta_{p2}) $ with respect to the shared ones. Examples of  triplets of  modes are shown in Fig.\ref{fig_PM0}, which plots the phase-matching surfaces $\Sigma_1 $ and $\Sigma_2$ in Eq. \eqref{Sigma0}, with shared modes at their intersections.
%,  as a function of  $ \lambda, \theta_x, \theta_y$, for symmetric pump tilts $\theta_{p1,p2} = \mp 1.2^\circ$. 
The green dots in  Fig.\ref{fig_PM0}b show the three entangled  modes at the degenerate wavelength, while  Figs. \ref{fig_PM0}c and \ref{fig_PM0}d illustrate the case of two conjugate wavelengths out of degeneracy,  showing two independent triplets of  modes,  labelled by dots and stars (notice that at any two conjugate wavelengths there are actually  4 independent triplets of modes). 

 If the emission frequency is not resolved, the various spectral components of the shared and coupled modes form in the far-field of the source three continuous branches at approximately  $\theta_x$ $\simeq  { \theta_{p1} + \theta_{p2}  \over 2} $ (shared modes) 
and $\theta_x \simeq  { \theta_{p1} + \theta_{p2}  \over 2} \pm   { (\theta_{p1} - \theta_{p2} )} $ (coupled modes). These are  shown by the numerical simulation in Fig.\ref{fig_simulazioni0}, where  shared and coupled modes appear as bright bands of hot spots against  the less intense background of the 2-mode fluorescence.  Notice that these simulations encompass a rather large bandwidth $  \Delta \lambda=260$ nm, so that the angular positions of high and low frequency spectral components split as predicted by Eq. \eqref{coupled_theta}. 
\par
Most important for our discussion, we notice that 
for a given finite tilt  $\pm  ( \theta_{p1} - \theta_{p2}) $ between the two pumps, the pattern of shared and coupled modes translates rigidly with  the angle of propagation ${ \theta_{p1} + \theta_{p2}  \over 2}$ of the career. As a consequence, the position of coupled modes  never superimpose to shared ones (stars never superimpose to dots in  Fig.\ref{fig_PM0}c, d. We will see in Sec. \ref{sec_BBO} a different phase-matching configuration, where such a superposition may take place, originating a transition to a 4-mode coupling. 

%%%%%%%%%%%%%%%%%%%%%%%%%%
\section{type I process in BBO: transition to a quadripartite entanglement} 
\label{sec_BBO}
This section studies a second configuration, where the two pump modes propagate inside a standard BBO crystal, forming in general  different angles with the optical axis. We shall see that the presence of strong walk-off effects enables a peculiar {\em resonance} condition,  with a transition to a 4-mode entangled state. %analogous to the one observed in nonlinear photonic crystals \cite{Jedr2018,Gatti2018}. 
\par 
%%%%%%%%%%% FIGURA SETUP fig_setup
\begin{figure}[ht]
\includegraphics[scale=0.53]{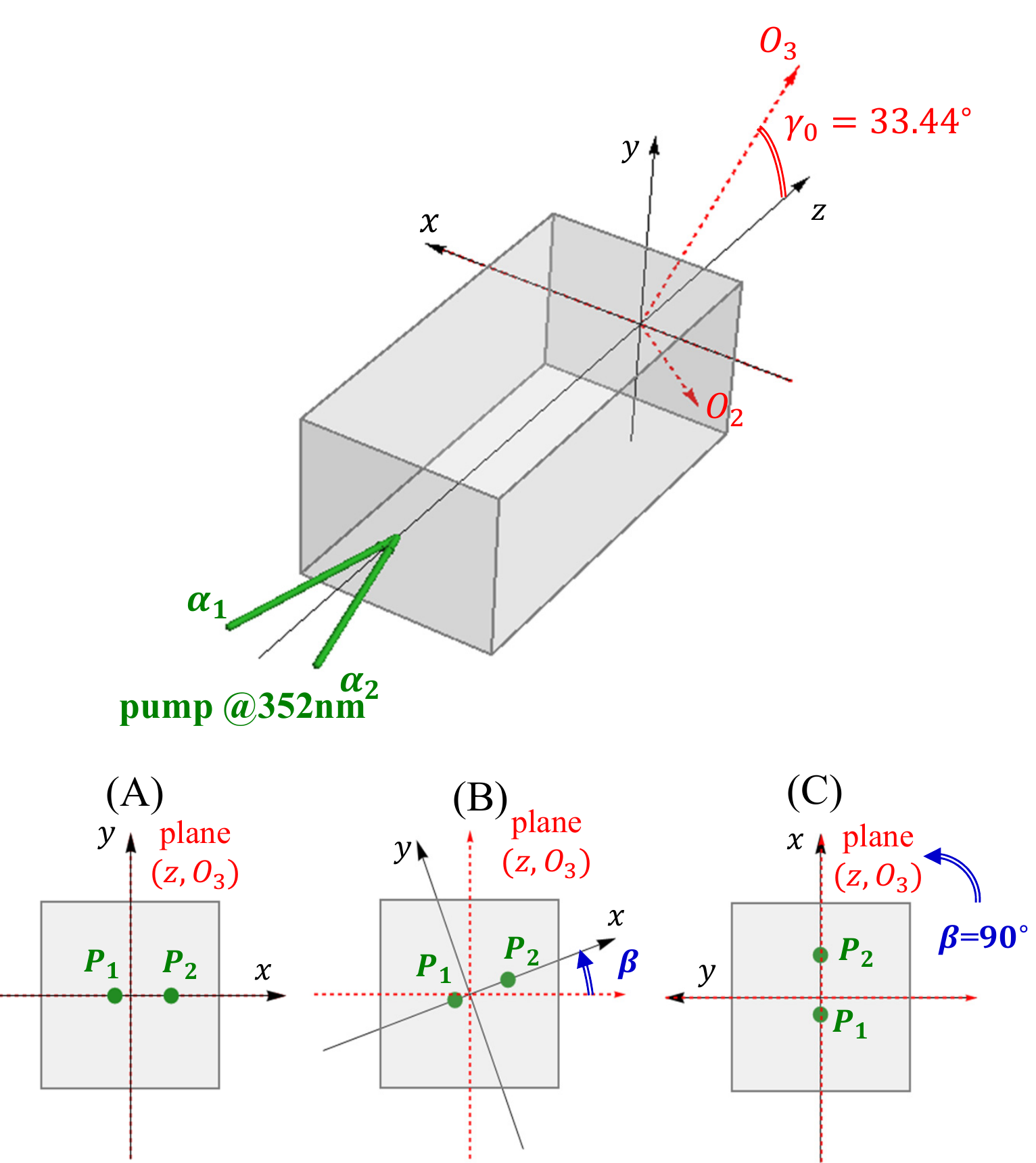}
\caption{ Geometry of the scheme, for the e$ \to $oo  process in  a BBO crystal, cut at $\gamma_0=33.44^\circ$.  $O_3$ is the optical axis.The  two pumps  propagate mainly along $z$, with a slight  tilt in the $x-$direction.  In the configuration  (A) the  pumps form roughly the same angle with the optical axis.  In  (B) and (C) the two pumps propagate at different angles with $O_3$, and have different wave-numbers.
}
\label{fig_setup1}
\end{figure}
%%%%%%%%%%%%%%%%%%%%%%%
We consider the same setup as in the experiment of Ref.\cite{Jedr2020}. The active material is a BBO crystal,  cut for the collinear type I process e$\to$oo from $\lambda_p = 352$nm to $\lambda_s=\lambda_i = 704$nm.  Fig.\ref{fig_setup1} shows the basic geometry:   the optical axis $O_3$ forms an angle $\gamma_0 \simeq 33.44^\circ$ with 
the mean propagation direction $z$. Unlike  the noncritical phase-matching of Fig.\ref {fig_setup0},   two pump modes slightly tilted with respect to $z$ experience in general   different refraction indices, because they propagate at different angles $\gamma_1$ and $\gamma_2$ with  the optical axis, and have different wave numbers $k_{p1}$ and $k_{p2}$, with   $k_{pj}= n_e (\omega_p, \gamma_j) \frac {\omega_p}{c}$. 
The difference  $ k_{p2} -k_{p1}$ 
 depends not only on the tilt angle, but also on the transverse direction of the tilt.  As we shall see in the following, the ability to tune this parameter enables the possibility to achieve the resonance associated to the 4-mode entanglement.   Fig.\ref{fig_setup1} A, B and C schematically depict the different geometries, where we associate  the  direction of the relative tilt between the pumps to the $x$ axis of a reference frame   $\{x,y,z\}$  which is allowed to rotate by an angle $\beta$
in the input facet of the crystal.  Notice that in practice  the various configurations are realized by    implementing a  $-\beta$ rotation of the  crystal around the $z$-axis  \cite{Jedr2020}. \\
% In agreement with the experimental implementationof Ref \cite{Jedr2020}, 
%In order to characterize the variation of the wave-number with the tilt angle, 
%we choose a fixed reference frame  $\{x,y,z\}$ in which the pumps are always tilted in the horizontal direction $x$, but we allow the crystal to rotate around the $z$ axis by an angle $\beta$, as shown in Fig.\ref{fig_setup1}A and B \footnote{This transverse rotation has not to be confused with the angle tuning}.
%  In the fixed reference frame
% the versors  associated to the direction of propagation of a generic  pump wave propagating at angle $\theta_p$ with the $z$ axis,    and  to the optical  axis $O_3$ are given by: 
%$$ \frac{ \vec k_{p}} {k_{p}}= \begin{pmatrix}  \sin{\theta_{p}} \\ 0 \\
% \cos {\theta_{p}} \end{pmatrix} 
%, \qquad 
%\etre = 
%\begin{pmatrix}  \sin \gamma_0  \sin  \beta  \\   \sin \gamma_0  \cos \beta \\
 %\cos \gamma_0 \end{pmatrix} 
%$$
%\nota{Le formule sopra sono gia' in appendice, posso non ripeterle, e lasciarle li', o toglierle dall'appendice, non so cosa e' piu' chiaro}
Then, as shown in App.\ref{app_shared} [see in particular  Eqs.\eqref{dkdq1}-\eqref{dgammadtheta}], for a given transverse  tilt   between the two pumps, the difference of their wave- numbers depends on the rotation $\beta$ according to the formula  (correct up to second order in the small angles $\theta_{p}$):
\begin{align}
 \frac{\Delta k_p}{\Delta Q_p}  = \frac{k_{p2} -k_{p1} } {Q_2 -Q_1}  %&= -\rho_{0}  \left. \frac{d \gamma}{d \theta_p }  \right|_{\bar \theta_p }   \\
& \simeq\left.  \rho_\gamma \left( \sin \beta  \cos \theta_p  \frac{ \sin \gamma_0 }{\sin \gamma } - \sin \theta_p \frac{ \cos \gamma_0 }{\sin \gamma }  \right)\right|_{\theta_p =   \frac{\theta_{p1} + \theta_{p2}} {2} }  
\label {dkdq}\\
&  \to \left\{ \begin{array}{lc}    \pm \rho_{\bar \gamma}   & \text{for } \beta = \pm  {\pi \over 2} \\   [0.4em]
\rho_0 \left( \sin \beta    - \frac{\theta_{p1} + \theta_{p2}} {2} \frac{1  }{\tg \gamma_0 } \right) & \text{for } |\beta |\ll  {\pi \over 2} \end{array} \right. 
\end{align}
 where 
$\rho_\gamma=-  \frac{1}{k_p}  \frac{d k_p}{d \gamma }  $  is the { \em walk-off angle } between  the wave-vector of the extraordinary wave and its Poynting vector, representing the direction of the energy flux  of the wave inside the medium \cite{BornWolf1999,Boeuf2000}. Here it is calculated at the angle $\bar \gamma$ formed by the carreer wave with the optical axis,  but we make a small error in taking it at the cut angle $\gamma_0$,  $ \rho_{\bar \gamma} \to \rho_{0}  \simeq 0.0744    \text{ rad} = 4.26^\circ$, according to the Sellmeier relations  in Ref. \cite{Kato86}. Clearly $\Delta k_p/\Delta Q_p$  is minimal  in the configuration labelled as A in   Fig.\ref{fig_setup1}  ($\beta= 0$),  and it  is maximal  in the configuration C  ( $\beta= \pm {\pi \over 2} )$, where it coincides with the walk-off angle between the  career wave and its Poynting vector. 
%Notice that in this last case the two pumps share the same principal plane,  and the Poynting vectors of both pumps points along the same $x$ direction as their relative tilt. 
%%%%%%%%%%%%%%%%%%%
\subsection{Shared-coupled modes and transition to resonance}
\label{sec_transition}
\begin{figure*}[ht]
\includegraphics[scale=0.65]{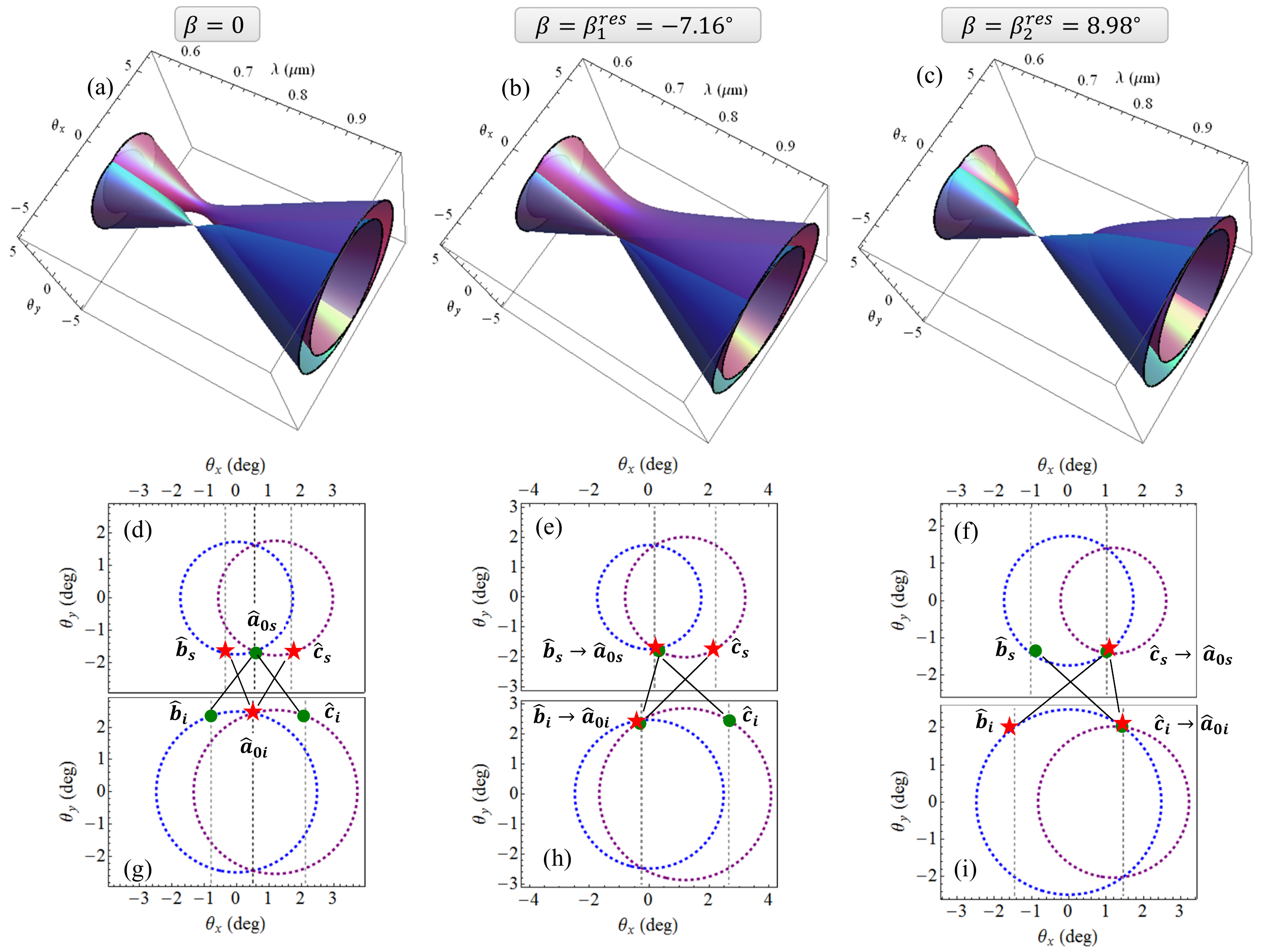}
\caption{BBO  doubly pumped at $352$nm.  $\theta_{p1} =0;  \theta_{p2}=1.2^\circ$.  (a,b,c) Surfaces in the $(\lambda,\theta_x,\theta_y)$ space defining phase-matching for the two pumps [see Eq.\eqref{Sigma0}],  calculated with the Sellmeier formulas in\cite{Kato86}. (d-i) Angular phase-matching curves at  the two conjugate wavelengths $\lambda_s=0.6 \mu$m (d,e,f)  and  $\lambda_i=0.85 \mu$m (g,h,i), illustrating the transition to resonance. The symbols show  the position of  shared and coupled modes. For $\beta=0$ (left column) two independent triplets  of entangled modes coexist  (dots and stars). At  $\beta_1^{\mathrm res} =-7.16^\circ$ (central column) and $\beta_2^{\mathrm res} =9.98^\circ$ (right column),  resonance is achieved. The two triplets of modes merge into four entangled modes}
\label{fig_PM1}
\end{figure*}

As in the former configuration of Sec.\ref{sec_PPLT}, each pump generates its own  branch of down-converted modes, laying  on the phase matching surfaces $\Sigma_1$ and $\Sigma_2$ defined by Eq.\eqref{Sigma0}.  Examples are shown in Fig.\ref{fig_PM1}, where the three columns correspond to three different  rotation angles  $\beta$.  At difference with the PPLT case of  Fig.\ref{fig_PM0}, we notice that now the surfaces $\Sigma_1$ and $\Sigma_2$  have quite different shapes, as discussed in Appendix A, and that their shape changes substantially with $\beta$. Actually, for the choice of parameters in this figure,  the crystal  rotation affects only the phase-matching  surface $\Sigma_2$,  which  changes from non-collinear (Fig.\ref{fig_PM0}b) for negative $\beta$,  to non-degenerate  for positive $\beta$ (Fig.\ref{fig_PM0}c).
%%%%%%%%%%% FIGURA  PM BBO fig_setup
%%%%%%%%%%%%%%%%%%%%%%%
\par
The geometrical intersections $\Sigma_1 \cap \Sigma_2$  determine the position of shared  modes $\w_0= (q_{0x}, q_{0y}, \Omega_0)$, each of them being  coupled to the  two  modes  $\w_{b} =   (Q_{1}-q_{0x}, -q_{0y}, -\Omega_0)$  and  $\w_{c} =   (Q_{2}-q_{0x}, -q_{0y}, -\Omega_0)$. Their Fourier coordinates are determined by Eq.\eqref{shared}, and are studied  in Appendix A [see Eqs. \eqref{q0x}-\eqref{q12}]. By translating these results into propagation angles around the $z$ axis, and neglecting infinitesimal terms $ \frac{\Dcoll (\Omega) }{k_p} \ll 1 $ \footnote{ $ \frac{\Dcoll (\Omega) }{k_p}  < 10^{-2}  $  for wavelengths in the whole interval  0.43-2.1 $\mu$m}
 we find that the angular positions of shared and coupled modes are given by 
\begin{align}
\theta_{0x}  (\Omega)& =  { \theta_{p1} + \theta_{p2}  \over 2}  + \frac{\Delta k_p}{\Delta Q_p} \frac{k_s (-\Omega) }{k_s (\Omega) }
\label{shared_theta1}\\
\theta_{b,c x}  (\Omega) &=  { \theta_{p1} + \theta_{p2}  \over 2}   \pm  { \theta_{p1} - \theta_{p2}  \over 2}  \frac{k_p}{k_s (\Omega) }
-  \frac{\Delta k_p}{\Delta Q_p} \, ,
\label{coupled_theta1}
\end{align}
while  in the y-direction
$\theta_{0y} (\Omega)= \theta_{b,c y} (\Omega)$. 
 In comparison with the noncritical phase-matching of Sec.\ref{sec_PPLT} [see  Eqs. \eqref{shared_theta} and \eqref{coupled_theta}] we  here see  the presence of   additional terms $\propto \frac{\Delta k_p}{\Delta Q_p}$, which have the effect of shifting the angular positions of shared and coupled modes in opposite directions. Thus,  by continuously varying this parameter,
%for proper values of the pump tilts 
one of the side modes may arrive to superimpose to  the central shared mode at the same frequency,  to which it was originally uncoupled, thus becoming itself shared.
As it  can be easily verified, the condition  $\theta_{b,c} (\Omega) =\theta_0 (\Omega)   $  takes place  for
\beq
\begin{aligned}
 \frac{\Delta k_p}{\Delta Q_p}   \left( 1 + \frac{\Dcoll(\Omega)}{k_p}  \right)&\simeq  \frac{\Delta k_p}{\Delta Q_p}  
=\begin{cases} 
 { \theta_{p1} - \theta_{p2}  \over 2}  \qquad & \theta_0 (\Omega)=\theta_{b} (\Omega)   \\
 { \theta_{p2} - \theta_{p1}  \over 2}  \qquad & \theta_0 (\Omega)=\theta_{c} (\Omega)  
\end{cases}
\label{resonance}
\end{aligned}
\eeq 
where again we neglected  $ \frac{\Dcoll (\Omega) }{k_p} \ll 1 $.   According to the results in Eq.\eqref{dkdq},  we notice that such conditions can be reached for any value  of  the  tilt angle  between the pumps smaller than the walk-off angle,  by  properly adjusting the transverse rotation of  the crystal. \par
We call the conditions in Eq.\eqref{resonance}  {\em resonances}, because of  their striking analogy with  the resonance that was observed in nonlinear photonic crystals \cite{Jedr2018,Gatti2018} by tilting the direction of a single pump wave inside the nonlinear grating. As for the NPC, at resonance two triplets of modes, originally uncoupled merge into a group a four modes, whose joint state is the quadripartite entangled state that will be described in Sec.\ref{sec_quadripartite}. Moreover, as demonstrated by the experiment of Ref.\cite{Jedr2020}, at resonance the parametric gain  of hot-spots undergoes a {\em Golden Ratio}  enhancement, again in perfect analogy with what observed in a hexagonally poled NPC \cite{Jedr2018}.
\par 
Fig.\ref{fig_PM1} provides an example of the  transition to resonance,  for the   two conjugate wavelengths $\lambda_s= 0.6 \mu$m and $\lambda_i= 0.85 \mu$m. In first column $\beta=0$, and the configuration is  analogue to the one studied in Sec.\ref{sec_PPLT}:   dots and the stars  correspond to two independent triplets of modes, which evolve according to the 3-mode propagation equation \eqref{3prop}, and whose  state is the tripartite entangled state described in Sec.\ref{sec_tripartite}.  In the second and third columns 
$\sin(\beta)=\pm   \frac{\theta_{p1} - \theta_{p2}}{2  \rho_0 } +  \frac{\theta_{p2} +  \theta_{p1}}{2 \tg \gamma_0 } $, respectively, corresponding to the two resonance conditions in Eq.\eqref{resonance}[See also Eq. \eqref{betares}].  At  $\beta=-7.16^\circ$,  all the shared modes merge with the left branch of coupled modes,  generated by pump 1:   $\azero, \azeroi \to \aunos, \auno$. At  $\beta=8.98^\circ$ the merging takes place between shared modes and the right branch of coupled modes, generated by pump 2:   $\azero, \azeroi \to \adues, \adue$. \\
Remarkably,  resonance is  achieved simultaneously for all shared-coupled modes in a   huge bandwidth around the degenerated wavelength. Indeed, 
even though the position of shared-coupled modes depends on the frequency,  the resonance condition does not: for any practical purpose,  the term $ {\Dcoll (\Omega) \over k_p }$ in Eq. \eqref{resonance} can be neglected because   $ {\Dcoll (\Omega) \over k_p } \simeq 
\frac{\Omega^2}{\Omega_B^2}$, where  $\Omega_B = \sqrt{ \frac{k_p}{k''_s} } \approx 2 \times  10^{16} \, \mathrm{s}^{-1}$. 
\par
Finally, the meaning of the resonance can be also appreciated by  reformulating Eq.\eqref{resonance}  in terms of Fourier modes, for which resonance is achieved  when   $\q_{0} (\Omega) = \q_{b,c} (\Omega) = \vec Q_{1,2} -\q_0 (-\Omega) $. This implies
\beq 
\q_{0} (\Omega) + \q_{0} (-\Omega) =\begin{cases}   \vec Q_{1}  \\
  \vec Q_{2}  
\end{cases} 
\label{qresonance}
\eeq 
where the upper (lower) condition corresponds to the upper (lower) condition in Eq.\eqref{resonance}.  Eq.\eqref{qresonance} is nothing else than the conservation of transverse momentum  involving  one  pump mode and the two shared modes: as a consequence, at  resonance,  shared modes at any  two conjugate wavelengths, otherwise uncoupled,   become populated by  photons pairs originating from the same pump mode.
We notice that   when one of the above conditions holds, not only shared modes superimpose to one branch of coupled modes, but also they become approximately collinear  in the x-direction  to one of the pump modes, as  can be easily checked from  Eq.\eqref{shared_theta1}  (see also  the examples in  Fig.\ref{fig_PM1}),

%%%%%%%%%%
\subsection{Quadripartite entanglement}
\label{sec_quadripartite}
This section studies the quadripartite entangled state generated at  resonance.  Let us concentrate for definiteness on the  upper resonance condition  in Eq.\eqref{resonance}, at which  two   shared modes  $\w_0$ 
and $\w_0^\prime$ 
 become coupled via the pump 1, thus satisfying  $\w_0  +\w_0^\prime = \Qone$.  The results for the other resonance can be obtained by exchanging $\bs,\bi \leftrightarrow \cs,\ci$ and $g_1 \leftrightarrow  g_2$. \\
Figs.\ref{fig_PM1}e) and h) show the schematics of the coupling in this case. Focusing on a specific pair of conjugate frequencies $ \pm \Omega_0$,   
%and  qy-coordinates$ \pm q_{0y}$, 
 labeled  by subscripts $s$ and $i$,  
the coupling involves the following  four modes: 
\begin{align}
& \text{shared modes} \begin{cases} \bs:= \As (\w_0 )  \\ \bi: = \As (\Qone -\w_0)    \end{cases}
& \text{coupled modes} \begin{cases} \cs:= \As (\Qtwo- \Qone + \w_0)  \\   \ci:= \As (\Qtwo -\w_0)  \end{cases}
\end{align}
Their evolution equations read
\begin{align}
\frac{d\bs}{dz} =&  \left[ g_1 \bi^\dagger + g_2 \ci^\dagger\right] e^{-i{\D (\w_0)} z}\nn \\
\frac{d\cs}{dz} =& \left[ g_2 \bi^\dagger  \right] e^{-i{\D(\w_0)} z}  \nn \\
\frac{d\bi }{dz} =&  \left[  g_1 \bs^\dagger + g_2 \cs^\dagger\right]e^{-i{\D(\w_0)} z}  \label{4prop}\\
\frac{d\ci }{dz} =&  \left[  g_2 \bs^\dagger  \right] e^{-i{\D(\w_0)} z} \nn
\end{align}
where the coupling coefficients $g_1 =\chi_1 \alpha_1$  and $g_2 = \chi_2\alpha_2$ are proportional to the pump amplitudes, but may also include a small effect due to  the different  nonlinear response of the medium in the two pump  directions.  The parameter $\D (\w_0)= \DD (\w_0, \Qone) = \DD (\w_0, \Qtwo)= \DD (\w_0, \Qtwo -\Qone)$ is the common value of the phase-mismatch, that must be assumed  small. 
%$| \D (\w_0) | \ll 1$
\par 
If one prefers the quantum state picture, 
then the evolution law of the  state associated to a quadruplet of modes  is easily found  in the simplest case of perfect phase-matching. For   $\D=0$, the propagation equations  \eqref{4prop} can be recast as $ \frac{d \hat O}{dz}  = \frac{1}{i \hbar} \left[ \hat {\cal P}, \hat O\right] $, where $\hat O = \bs ...\ci $, and the "momentum" operator is $ \hat  {\cal P}   = {-i \hbar} \left[ g_1 \bs^\dagger \bi^\dagger +
 g_2 \left(  \bs^\dagger \ci^\dagger + \cs^\dagger \bi^\dagger \right)   - \text{h.c} \right] $. Then the state evolves according to 
\begin{align}
|\psi\rangle_{out}  = & e^{\frac{i}{\hbar}  \hat {\cal P} z}  \, |\psi\rangle_{in} \nn \\
=&
e^{ \left[ g_1 \bs^\dagger \bi^\dagger 
+ g_2 \left(  \bs^\dagger \ci^\dagger + \cs^\dagger \bi^\dagger \right)  -\text{h.c} 
 \right] z }  |\psi\rangle_{in} 
\label{statePDC} \\
 \underset{\gbar z \to 0}  { \to} &
|\psi\rangle_{in}  + 
z \left[ g_1 \bs^\dagger \bi^\dagger + g_2  (  \bs^\dagger \ci^\dagger + \cs^\dagger \bi^\dagger ) 
 \right] \, |\psi\rangle_{in}  
\label{stateSPDC}
\end{align}
where in writing Eq.\eqref{statePDC} we assumed some form of discretization of Fourier modes (details not relevant to our discussion).  Eqs. \eqref{statePDC} or \eqref{stateSPDC}
 show the two-photon processes occurring in the quadruplet of modes:  a photon pair may be down-converted from pump 1, with  probability amplitude $g_1 \propto \alpha_1$, and appear in modes $\bs,\bi$. Alternatively, paired photons can be generated from pump 2 (probability amplitude $g_2 \propto \alpha_2$)
and appear in one of the two couples  $(\bs, \ci)$ or  $(\cs, \bi)$.  When  one of the two pumps is absent, the state reduces to a product of   bipartite EPR states generated by each pump;  for example, for $g_1=0$  the equations show the contribution of two independent couples of entangled modes over the many couples  generated by down-conversion from pump 2.
\par
\begin{figure}[ht]
\includegraphics[scale=0.55]{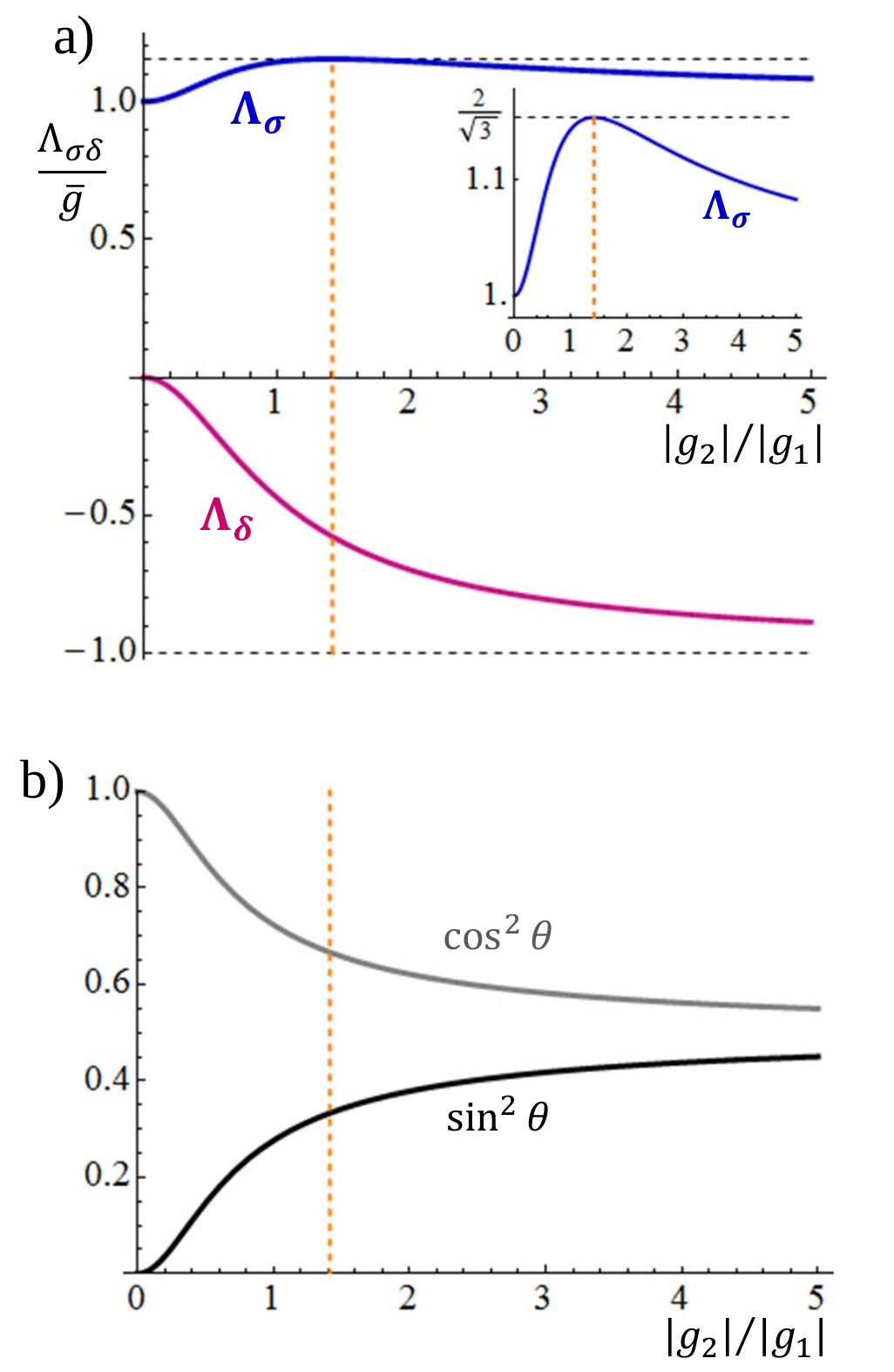}
\caption{a) Squeeze eigenvalues $\Lambda_\sigma$ and $\Lambda_\delta$ in Eq.\eqref{Lambda}, normalized to the reference squeeze parameter  $\bar g  $ of   a single pump carrying all the energy. The inset shows the maximum of  $\Lambda_\sigma$, occurring at $ |g_2| = \sqrt {2} |g_1|$. b) Mixing coefficients of the unitary  transformation in Eq.\eqref{U}}
\label{fig_Lambda}
\end{figure}

 
As for  any  multipartite Gaussian entangled state \cite{Braunstein2005},  the  quadripartite  state in Eq.\eqref{statePDC} can be decomposed into 4 single-mode  squeezed states mixed by passive linear transformations. In our case, we prefer a decomposition into a pair of of bipartite EPR states  (each  of them can be in turn thought of as the balanced  interference of two squeezed states). This decomposition    is accomplished by 
%The propagation equations \eqref{4prop} can be solved by a introducing 
the following linear  transformation acting  separately on the signal and idler modes
\begin{align}
&\begin{pmatrix}
\hat b_j \\
\hat c_j 
\end{pmatrix}  = 
\matr{U}
\begin{pmatrix}
 \sigmak_j \\   \deltak_j
\end{pmatrix}
\quad (j=s,i) \, ,
\label{canonical0}
\\
%where the  2x2 unitary  matrix   $\matr{U} $  has the form: 
&\matr{U}  = 
\begin{pmatrix}
e^{i \frac{\phi_1}{2}} & 0 \\
0 &  e^{i \frac{\phi_2}{2}} e^{-i {\phi_-}} 
\end{pmatrix} 
\cdot 
\begin{pmatrix}
\cos \theta & \sin \theta \\
-\sin \theta & \cos \theta 
\end{pmatrix} 
\label{U} 
\end{align}
where the  mixing coefficients $\cos \theta $ and $\sin \theta$  depend on the ratio between the two pump intensities, as described 
by Eq.\eqref{Upar} and shown by Fig.\ref{fig_Lambda}b. 
%%%%%%%%%%% FIGURA  PM BBO fig_setup
%%%%%%%%%%%%%%%%%%%%%%%
 Under this transformation the equations \eqref{4prop} decouples  into two independent standard parametric processes of the form 
\begin{align}
\frac{d }{dz}\deltas(z) &= \Lambda_\delta \,  \deltai^\dagger (z) e^{-i \Dbar z} \nn \\
\frac{d}{dz}\deltai(z)  &= \Lambda_\delta  \,\deltas^\dagger (z) e^{-i \Dbar z} 
\label{deltaprop}
\end{align}
and 
\begin{align}
\frac{d }{dz} \sigmas(z) &= \Lambda_\sigma \, \sigmai^\dagger (z) e^{-i \Dbar z} \nn \\
\frac{d }{dz} \sigmai (z) &= \Lambda_\sigma  \, \sigmas^\dagger (z) e^{-i \Dbar z} 
\label{sigmaprop}
\end{align}
Both the squeeze parameters $\Lambda_{\sigma}$,  $\Lambda_{ \delta}$ and the mixing coefficients of the unitary $\matr{U}$ depend only on the ratio 
\beq
\rho = \frac{|g_2| }{ |g_1|} \propto  \frac{|\alpha_2| } {|\alpha_1|} , 
\eeq
 according to the following equations: 
\begin{align}
\Lambda_{\sigma \delta} & =| g_1 | f_{\pm} (\rho) = \frac{ \gbar }{\sqrt{1+ \rho^2}} f_{\pm} (\rho) 
%  \frac{1 \pm \sqrt{1 + 4\rho^2}}{2}  \\
%&=\frac{ \gbar }{\sqrt{1+ \rho^2}}  \frac{1 \pm \sqrt{1 + 4\rho^2}}{2} 
\label{Lambda} \\
\cos \theta  &=\frac{ f_+ (\rho) }{\sqrt{ \rho^2 + f_+^2 (\rho) }  } ; \quad 
\sin \theta = -\frac{ \rho }{\sqrt{ \rho^2 + f_+^2 (\rho) }  } 
\label{Upar}\\
f_\pm (\rho) &= \frac{1 \pm \sqrt{1 + 4 \rho^2}}{2}.  \nn
\label{Upar2}
\end{align} 
In these formulas $\gbar= \sqrt{|g_1|^2 + |g_2|^2} $ is the reference squeeze parameter, corresponding to standard PDC pumped by a single beam carrying the total energy of the two modes  (apart from minor corrections arising from different nonlinear coefficients). 
\\
Under the same transformation the output state reduces to the product of two independent EPR states in modes $\sigmak_j$ and $\deltak_j$, 
$ |\psi\rangle_{out}  \to
e^{ \left[ \Lambda_\sigma\sigmak_s^\dagger \sigmak_i^\dagger  -\text{h.c} 
 \right] z }   
e^{ \left[ \Lambda_\delta \deltak_s^\dagger \deltak_i^\dagger  -\text{h.c} 
 \right] z }    |\psi\rangle_{in} $. 
Figure \ref{fig_unfolding1} shows the unfolding of the state: the 4-mode entangled state generated at resonance is formally equivalent to:  i) {\em two} nonlinear processes, each generating an EPR pair with squeeze parameters $\Lambda_{\sigma }$ and $\Lambda_\delta$;  followed by ii) a beam splitter with transmission and reflection coefficients $t=\cos \theta$ and $r=\sin \theta$, respectively, which mixes the two EPR pairs, and  iii) phase rotations in the two output arms,  by $\frac{\phi_1} {2} $ 
and  $ \frac{\phi_2} {2}- \phi_-$, respectively. We remark that  the  splitting ratio $r/t$ of the  beam-splitter  can be varied by modulating the  pump intensities [Eq.\eqref{Upar}],   which means  the scheme is potentially able to produce any arbitrary mixing of a pair of EPR states, offering in this way   the  possibility of engineering the 4-mode entanglement. 
\par
A quantitative characterization of the  entanglement of this state is outside the scopes of this work and will be performed elsewhere \cite{Gatti2020c}. We simply notice that all the modes interact in a linear chain,  shown e.g. by the scheme in  Fig. \ref{fig_PM1}e,h. Their genuine quadripartite entanglement can be demonstrated by noticing  that there exist four independent combinations of mode quadrature operators whose variances vanish in the limit of large squeezing, violating thus any bound imposed by separability. Precisely,  by defining the quadrature operators of modes $\hat b_j$ relative to the phase of pump 1, i.e,  $ \hat X_{b_j} = 
\hat b_j e^ {-i \frac{\phi_1}{2} }+ \hat b_j^\dagger e^ {+i \frac{\phi_1}{2}} $,  defining  those of modes $c_j$ as $ \hat X_{c_j} = 
\hat c_j e^ {-i \frac{\phi_2}{2}} e^{i \phim}  + \hat c_j^\dagger e^ {+i \frac{\phi_2}{2}} e^{-i \phim}  $,  by using the inverse of transformation \eqref{U} and the standard properties of EPR states, it can be shown that 
\beq
\begin{aligned}
 \hat f_{\scriptscriptstyle\MakeUppercase{\romannumeral 1}    } &=\cos \theta ( \hat X_{b_s} -\hat X_{b_i}) -\sin \theta  \, ( \hat X_{c_s} -\hat X_{c_i})= \sqrt{2} e^{- \Lambda_\sigma z} 
\hat X_{\scriptscriptstyle\MakeUppercase{\romannumeral 1}    } (0) \\
 \hat f_{\scriptscriptstyle\MakeUppercase{\romannumeral 2}    } &=\sin \theta ( \hat X_{b_s} +\hat X_{b_i})  +\cos \theta  \, ( \hat X_{c_s} +\hat X_{c_i})= \sqrt{2} e^{- |\Lambda_\delta| z} 
\hat X_{\scriptscriptstyle\MakeUppercase{\romannumeral 2}    }(0) , 
\end{aligned}
\eeq
where $\hat X_{\scriptscriptstyle\MakeUppercase{\romannumeral 1}    } (0)$ and $\hat X_{\scriptscriptstyle\MakeUppercase{\romannumeral 2}    } (0)$ are independent input operators that can be taken in the vacuum state. At the same time for the orthogonal quadratures $\hat Y_{\alpha}$, such that $\left[ \hat X_\alpha, \hat Y_\beta\right] = 2 i \delta_{\alpha,\beta} $, $(\alpha, \beta = b_s ..c_i)$, one can show that 
\beq
\begin{aligned} 
 \hat f_{\scriptscriptstyle\MakeUppercase{\romannumeral 3}    } &=\cos \theta ( \hat Y_{b_s} +\hat Y_{b_i}) -\sin \theta  \, ( \hat Y_{c_s} +\hat Y_{c_i})= \sqrt{2} e^{- \Lambda_\sigma z} 
\hat Y_{\scriptscriptstyle\MakeUppercase{\romannumeral 3}    } (0)  \\
 \hat f_{\scriptscriptstyle\MakeUppercase{\romannumeral 4}    } &=\sin \theta ( \hat Y_{b_s} -\hat Y_{b_i})  +\cos \theta  \, ( \hat Y_{c_s} -\hat Y_{c_i})= \sqrt{2} e^{- |\Lambda_\delta| z} \hat Y_{\scriptscriptstyle\MakeUppercase{\romannumeral 4}    } (0) 
\end{aligned} 
\eeq
where  again $\hat Y_{\scriptscriptstyle\MakeUppercase{\romannumeral 3}    } (0)$ and $\hat Y_{\scriptscriptstyle\MakeUppercase{\romannumeral 4}    }(0)$ are independent vacuum  operators.  
The  observables $\hat f_{\scriptscriptstyle\MakeUppercase{\romannumeral 1}    }, ...\hat f_{\scriptscriptstyle\MakeUppercase{\romannumeral 4}    } $ commute pairwise, so that in general there is no lower  bound for their variances. However,  in the same spirit of \cite{Simon2000, Furosawa2003}, it is possible to formulate bounds that must be satisfied  by  states separable with respect to any specific bipartition, which are violated when the  gain $\gbar z $ is large enough, provided that  both $|\alpha_1| \ne 0 $ and $ |\alpha_2| \ne 0 $   \cite{Gatti2020c}. 
\par
Interestingly, the bigger squeeze eigenvalue is always slightly  larger than $\gbar$, and  presents a maximum at $|g_2| = \sqrt{2} |g_1|$, i.e. when the pump 2 is approximately twice as intense as pump 1, where
 $ \Lambda_\sigma =  \frac{2}{\sqrt 3} \gbar \simeq 1.15 \gbar$.    This means that at resonance the  doubly pumped scheme achieves a larger  amount of squeezing/gain  in the auxiliary modes $\sigmak_j $ with respect to a standard single-pump scheme, at the same level of injected energy. In this way squeezing/entanglement is concentrated  in specific modes. 
\par 
Finally, we  notice that  for $\rho=1 $, i.e. when  the two pump intensities are balanced, the  squeeze eigenvalues reduce to $ \Lambda_\sigma= |g_1| \Phi$
% \frac{ \gbar }{\sqrt{2}} \Phi$ 
and 
$ \Lambda_\delta=-  \frac{|g_1| }{\Phi} $
% \frac{ \gbar }{\sqrt{2}} \frac{1}{\Phi}$, 
where $\Phi= \frac{1 + \sqrt{ 5}}{2}$ is the  Golden Ratio: in this case the doubly pumped PDC scheme  realizes a complete analogy
% (a part from a $\sqrt{2}$ factor)
 with the { "Golden Ratio Entanglement"} demonstrated in   a hexagonally poled photonic crystal \cite{Gatti2018} with a single pump. An interesting comparison is also   with  the quadripartite entanglement generated  in a {\em doubly} pumped nonlinear photonic crystal  \cite{Gatti2020}: in this case  the squeeze eigenvalues are  controlled not only by the relative intensity but also  by the relative {\em phase} of the two pumps, which allows to acces a larger variety of states. . 
%%%
\subsection{The resonance and the Poynting vectors} 
\begin{figure}[ht]
\includegraphics[scale=0.6]{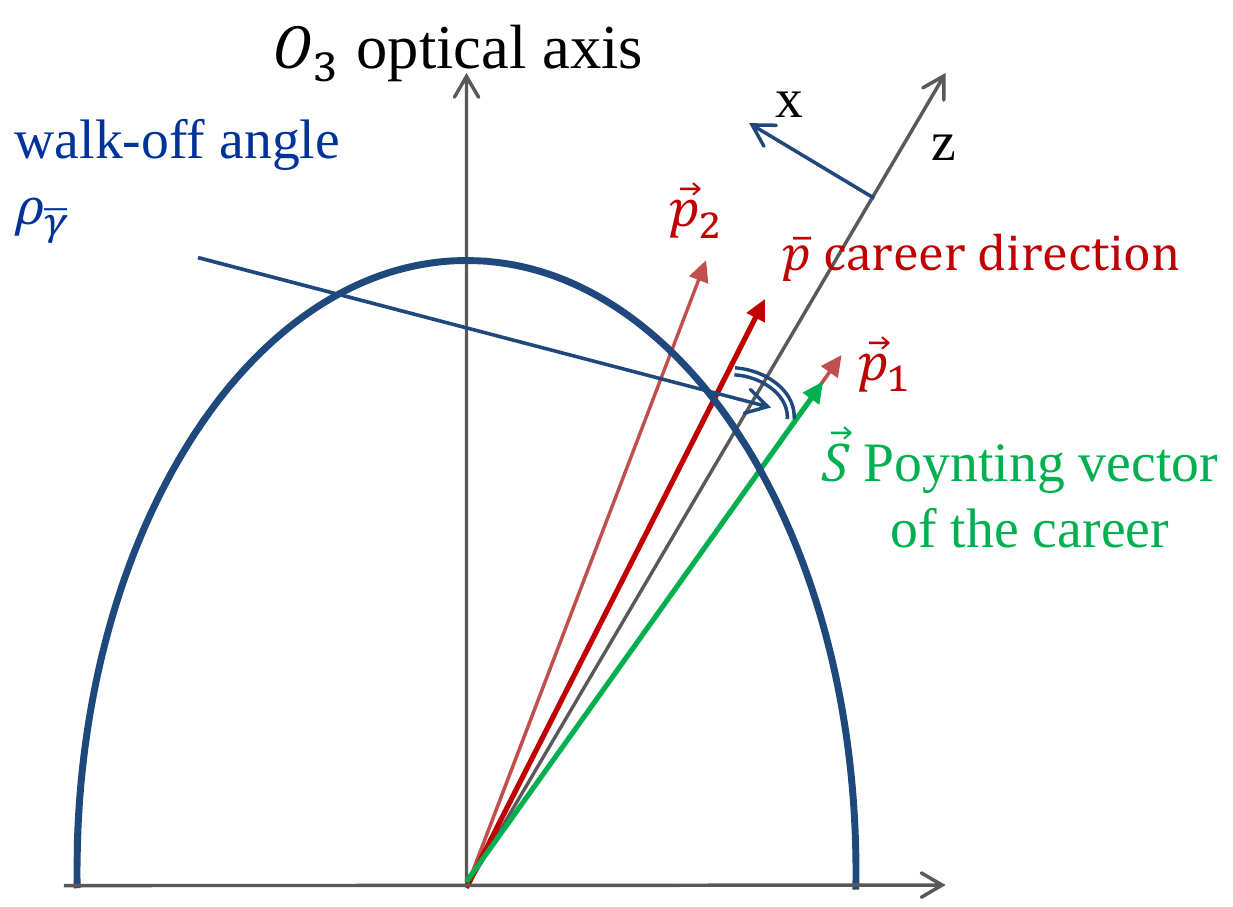}
\caption{ Illustration of the resonance in the case  $\beta = {\pi \over 2}$. The blue thick line represents the ellipsoid of the refraction indexes, $\vec{p}_1$, $\vec{p_2}$ and  $  \bar p $ show the directions of propagation of pump 1, pump2, and of the pump career respectively, while $\vec S $ is the propagation direction of its  Poynting vector. At resonance,  it superimposes to the direction of propagation of pump 1}
\label{fig_interpretation}
\end{figure}

The resonance, as we called the transition from 3 to 4-mode entanglement, admits an interesting interpretation in terms of a superposition between the Poynting vector of the pump career, representing the mean direction of propagation of the energy flux, and one of the pump modes. \\
 This interpretation is particularly evident in the configuration  C of Fig.\ref{fig_setup1},  and is illustrated in Fig.\ref{fig_interpretation}. 
 In this case, the problem  becomes 2-dimensional because the  pump modes share the same principal plane,  which includes  the optical axis $O_3$.  The wave-vector  of the pump career lies at an angle $\frac{\theta_{p2} +\theta_{p1}}{2}$  from the z-axis, and 
 its Poynting vector  walks-off  in the principal plane  by an amount $\rho_{\bar \gamma}$, away from the optical axis (BBO is a negative uniaxial crystal), i.e. it forms an angle $ \theta_{\bar S}= \frac{\theta_{p2} +\theta_{p1}}{2} -\rho_{\bar \gamma} $ with the $z$-axis.   For  $\beta = \frac{\pi}{2}$, 
the resonance conditions,  described by Eq.\eqref{resonance}  and  Eq.\eqref{dkdq},   reduce to
$
\rho_{\bar \gamma} =\mp  \frac{\theta_{p2} - \theta_{p1}} {2}  
$. For our choice of parameters $\theta_{p2} - \theta_{p1} >0$,  and only the lower condition   can be satisfied, leading to  $  \theta_{\bar S}= \theta_{p1}. $
For $\beta = -\frac{\pi}{2}$  the roles of p1  and p2  are exchanged (the x axis is reversed), leading to 
\beq
 \theta_{\bar S}=\begin{cases}
 \theta_{p2} &\quad  \text{for } \beta = - {\pi \over 2} \\
 \theta_{p1} &\quad  \text{for } \beta = + {\pi \over 2} 
\end{cases} 
\eeq
i.e. to the result that the  resonance condition exactly corresponds to the superposition between the direction of propagations of   the Poynting vector of the career and one of the pump modes. \\
The general case is slightly more involved, because of the full 3-dimensional geometry of the problem. It becomes quite clear when the pump tilts are symmetric, i.e. the pump career propagates along z.  Then its Poynting vector  points as in Fig.\ref{fig_interpretation} and  the resonance condition becomes $\rho_0 \sin \beta = \mp  \frac{\theta_{p2} - \theta_{p1}} {2} $.  For $\beta=0$  the plane $(\vec p_2, \vec p_1)$ is perpendicular to the plane of the figure, and there is no possibility of superposition. . 
For $\beta \ne 0$,   the transverse component of the Poynting vector in the $x$ direction of the tilt can superimpose to one of the pump modes, allowing thus a resonance. 

%%%%%%%%%%%%%%%%%%%%%
\section{ Conclusions}
This work has analysed two doubly pumped schemes of parametric down-conversion, in realistic experimental configurations, which exploit standard and commercially available nonlinear media.   It has highlighted a  stringent analogy with the phenomena predicted and observed   in 2-dimensional nonlinear photonic crystals, by using simpler sources which do not need lengthy poling procedures, and offering in addition the possibility of reconfigurating some properties of the state by a simple modulation of the classical laser beam driving the process. 
\par
In the non-critical phase-matching case of the PPLT our analytical results, complemented by numerical simulation, may constitute a proposal for future experimental implementations. In our opinion the main outcome here concerns the possibility of implementing an arbitrary beam-splitter on one of two parties of the EPR state  generated by standard parametric down-conversion by acting on the spatial structure of the classical laser beam rather then on the fragile quantum state. 
\par 
The BBO case has already found an experimental demonstration for what concerns the classical properties of the process \cite{Jedr2020}. For the quantum properties, the highlight result is the  possibility of directly  generating quadripartite entangled states, and of modulating their properties by acting on the intensities of the two pump modes. This possibility is enabled by the walk-off effects present in such an  anistropic material, in a way that  is in our opinion highly nontrivial. 
In particular, the 4-mode entanglement can be realized at any small tilt angles between the pumps (namely provided that the tilt angle is smaller than the walk-off angle in the central direction of light propagation).  We offered also an interpretation of the   resonance, as we called the transition from 3- to 4-mode entanglement, 
 in terms of a superposition between the career  Poynting vector, which identifies the direction of propagation of the energy flux, with either one pump mode or the other. 
\appendix
\section{Analytical calculations in paraxial approximation}
\label{app_shared}
This Appendix summarizes some  analytic  results, derived by using  the paraxial approximation. The dependence on the frequency $\Omega$ is maintained  till the very end, because we are interested in large emission bandwidths. 
 Precisely,  the z-component of the signal wave-vector is approximated as: 
\beq
k_{sz} (\q,\Omega)= \sqrt{k_s^2(\Omega) -q^2}  \to  k_s (\Omega) -\frac{q^2}{2k_s (\Omega)} 
\label{ksz}
\eeq
valid for $ q \ll k_s (\Omega)$ (small angles around the $z$). The wave-number $k_s (\Omega)$ does not depend on the propagation  direction  because a) in the PPLT case the down-converted light propagates close to $\pi \over 2$ (and the material has a very small birefringence), and b) in  the BBO case the signal is an ordinary wave.  For the extraordinary pump waves: 
\beq
k_{pzj} = \sqrt{k_{pj}^2-Q_j^2}  \simeq k_{pj}  -\frac{Q_j^2}{2k_{p}}  \quad (j=1,2) 
\label{kpz}
\eeq
where $k_{p}= n_e (\omega_p, \gamma_0) \frac {\omega_p}{c}$, and:  a)   in the PPLT case $k_{pj}=k_p$; b)  in the BBO case $k_{pj}= n_e (\omega_p, \gamma_j) \frac {\omega_p}{c}$ depends  on the angle $\gamma_j$ formed by the wave with the optical axis $O_3$. Let us consider   the geometry  in Fig.\ref{fig_setup1}, where  the transverse tilt of the pump takes place along the x-direction, inclined  at an angle $\beta$ in the input facet of the crystal.  In the reference frame 
$(x',y',z) $ parallel to the facets of the crystal  [not to be confused with the crystalline reference frame $(O_1,O_2,O_3) $)], 
 the versors  associated with the direction of propagation of a generic  pump wave   and  with the optical  axis $O_3$ are respectively:
\beq
 \frac{ \vec k_{p}} {k_{p}}= \begin{pmatrix}  \sin{\theta_{p}} \cos \beta   \\   \sin{\theta_{p}} \sin \beta   \\
 \cos {\theta_{p}} \end{pmatrix} 
, \qquad 
\etre = 
\begin{pmatrix}  0 \\   \sin \gamma_0  \\
 \cos \gamma_0 \end{pmatrix} 
\label{e3}
\eeq
The angle formed by  the pump with the optical axis is thus determined by 
\beq
\cos{\gamma} =  \frac{ \vec k_{p} \cdot \etre} {k_{p}} =  \cos {\theta_{p}}\cos \gamma_0 +  \sin \theta_{p} \sin{\gamma_0} \sin{\beta} .
\label{gamma}
\eeq
For small pump tilts,  the variation of $\gamma$ with  $\theta_p$ is minimal  for $\beta= 0$ (as in Fig.\ref{fig_setup1}A), where $\cos \gamma \simeq \cos \gamma_0 (1-\frac{\theta_p^2}{2})$, while it is maximal for  $\beta=\pm 90^\circ$, where $\gamma= \gamma_0  \mp \theta_p$.
%%%%%%%%%%%%%%
\renewcommand{\thesubsubsection}{\thesection\Roman{subsubsection}}
%\renewcommand{\thesubsubsection}{\Roman{subsubsection}}
%\subsubsection{ Phase matching surfaces} 
% \label{app_PM}
\\

 {\bf \small Phase matching surfaces} \\
By inserting the approximated expressions \eqref{ksz} and \eqref{kpz} into the definition of the phase matching function in Eq. \eqref{DD}, and performing some long but  simple algebra, the equation for the phase matching surfaces $\Sigma_1$ and $\Sigma_2$ defined in Eq. \eqref{Sigma0} can be obtained as: 
\begin{align}
&\left| \q - \vec Q_j \frac{k_s (\Omega) }{ k_s (\Omega) + k_s (-\Omega) }  \right|^2  = F_j  (\Omega)  & (j=1,2) ,
 \label{PMeq} \\
&
F_j  (\Omega) = \bar k (\Omega)  
\left[   \Dcoll (\Omega) -(k_{pj} - k_p) +\frac{ \vec Q_{j}^2}{k_p} 
% \left[ 1- \frac{k_p}{k_s(\Om) + k_s(-\Om)}\right]   \right]  &
\, \frac{ \Dcoll (\Omega) -G_z  }{ k_p + \Dcoll (\Om) -G_z}  \right]  &
\label{PMrad}
\end{align} 
where $ \bar k (\Omega) =\frac{2 k_s (\Omega) k_s (-\Omega) }{k_s (\Omega) + k_s (-\Omega) } $, and 
$ \Dcoll (\Omega) = k_s (\Omega) + k_s (-\Omega) -k_p +G_z $ is the collinear phase-mismatch function.  In Eq.\eqref{PMrad}  one must take:  a)  $G_z \ne 0$  and  $ k_{pj} - k_p=0$ for the PPLT;  b) $G_z= 0 $ for the BBO (poling is absent). 
For the frequencies such that $F_j  (\Omega)  >0$, Eq. \eqref{PMeq}  represents a family of circumferences,  centered around $ q_{jx}^c (\Omega) = 
 Q_j  \frac{ k_s (\Omega) }{ k_s (\Omega) + k_s (-\Omega) } \simeq\theta_{pj} k_s(\Omega) $. Thus
the angular coordinate of the center   is  approximately  $ q_{jx}^c (\Omega) /k_s(\Omega)  \simeq   \theta_{pj}$. As expected, the two emission branches are  conical surfaces  roughly collinear with each pump, examples being shown in figures \ref{fig_PM0} and \ref{fig_PM1}.  The shape of each surface depends on the value of 
$F_j  (\Omega=0)  $ in the standard way, i.e. it is a  open tube  for  $F_j  (0)  >0 $, which  collapses to  a "hourglass"  for $F_j  (0) =0 $, while it presents two separate branches for  $F_j  (0)  <0 $. 
Notice that in the PPLT case the shape changes slowly with the tilt angle,so that the two phase-matching branches look very similar (see Fig. \ref{fig_PM0}),  while  in the BBO case it has a much faster variation due to the term $k_{pj} - k_p$, so that in general $\Sigma_1$ and $\Sigma_2$ look quite different  (see Fig. \ref{fig_PM1}). 
%%%%%%%%%%%%%%%%%%%%%
%\subsection{Shared and coupled modes.}
\\

\par
{\bf \small Shared and coupled modes.}\\
The Fourier coordinates of  shared modes and of their coupled ones is determined  by Eq.\eqref{shared}. By  imposing the shared mode condition 
$\DD  (\w_0; \Qone)= \DD  (\w_0; \Qtwo)  $,  using again Eqs \eqref{ksz} and \eqref{kpz}, and reordering the various terms, one obtains  the following condition on the 
x-component of the  wave- vector: 
\beq
q_{0x}  (\Omega) =  { Q_1 + Q_2 \over 2}  \left( 1- \frac{k_s (-\Omega)}{k_p} \right) + \frac{\Delta k_p}{\Delta Q_p} k_s (-\Omega)
\label{q0x}
\eeq 
where 
\beq 
 \frac{\Delta k_p}{\Delta Q_p}=\frac{k_{p2} -k_{p1}}{Q_2 -Q_1}
\label{dkdqA}
\eeq 
measures the rate of variation  of the pump wave-numbers  with their transverse tilts. Such a term is absent in the PPLT scheme, but plays a crucial role in the BBO case because of the strong birefringence of the material. 
The y-component of the wave-vector is obtained by requiring that phase matching is satisfied, i.e. that  $\DD  (\w_0; \Qone)=\DD  (\w_0; \Qtwo)=0$.  Using Eq. \eqref{PMeq},  one has
\begin{align}
&q_{0y}  (\Omega) = \pm \sqrt{ F_j (\Omega) -[q_{0x} -q_x^c]^2}  %&  \text{for  }  F_j (\Omega) -[q_{0x} -q_x^c]^2 \ge 0
\label{q0y}
\end{align}
for  $ F_j (\Omega) -[q_{0x} -q_x^c]^2 \ge 0$ , i.e. provided that the intersection between $\Sigma_1$ and $\Sigma_2$  exists.  The $\pm$ signs  correspond to the two possible intersection points of two circumferences. \\
The modes coupled to each shared mode have equation $ \q_b (\Omega) = \Qone -  \q_0 (-\Omega)$  (via pump 1) and 
$ \q_c (\Omega) = \Qtwo-  \q_0 (-\Omega)$  (via pump 2). At a given frequency $\Omega$,  their transverse coordinates are: 
\beq
\begin{aligned}
q_{b,c \,x}  (\Omega) &= Q_{1,2} -  { Q_1 + Q_2 \over 2}  \left( 1- \frac{k_s (\Omega)}{k_p} \right) - \frac{\Delta k_p}{\Delta Q_p} k_s (\Omega) \\
q_{b,c \,y }  (\Omega) & = -q_{0y}  (-\Omega)  = \pm q_{0y}  (\Omega) 
\end{aligned} 
\label{q12}
\eeq
where the last equality follows from the symmetry  of equations \eqref{q0y} and \eqref{PMrad}  with respect to the exchange $\Omega \to -\Omega$ . 
%\subsection{Resonance condition}
\\

\par
{\bf \small  The resonance.}\\
We use here the resonance condition in Eq. \eqref{qresonance} $\q_0 (\Omega) + \q_0 (-\Omega) =  \vec Q_{1,2} $. This equation  can be always satisfied for the y-coordinate,  since $F_j (\Omega)$  in  Eq.\eqref{q0y} is an even function of  $\Omega $, so that  one can choose $q_{0y}(-\Omega) = - q_{0y}(\Omega)$. For the x-coordinate, using Eq.\eqref{q0x}, it requires  that
\begin{align}
Q_1 + Q_2 &+  [ k_s (\Omega + k_s(-\Omega) ]\left( \frac{\Delta k_p}{\Delta Q_p} -\frac{Q_1 + Q_2} {2k_p}\right) = Q_{1,2} \label{res0}\\
\to \frac{\Delta k_p}{\Delta Q_p} &=% \frac{Q_1 + Q_2} {2k_p} - \frac{Q_{2,1} }{k_p}\frac{k_p}{k_s (\Omega + k_s(-\Omega) }  
\frac{\theta_{p1} + \theta_{p2}} {2} - \theta_{p2,p1} \frac{k_p}{k_s (\Omega + k_s(-\Omega) }  \nn \\
& =  \pm 
\frac{\theta_{p1} - \theta_{p2}} {2} +\theta_{p2,p1} \frac{\Dcoll (\Omega)-G_z }{k_p  + \Dcoll (\Omega)-G_z }  
\label{res1}
\end{align} 
where, as usual, we approximated $\theta_{pj} \simeq  \frac{Q_j}{k_p}$, and we used the identity  $k_p = k_s (\Omega) + k_s(-\Omega)  - \Dcoll (\Omega) + G_z  $. First of all, we notice that  the second term at r.h.s. of  Eq.\eqref{res1} is a  very small correction, because  $|G_z - \Dcoll (\Omega)| \ll k_p$. Thus, Eq.\eqref{res1}  cannot be satisfied when  $ \Delta k_p =0$ because it  would require   $ | \theta_{p1} - \theta_{p2} | \ll |\theta_{p2,p1}|$  (in practice that the pump modes  are collinear). Therefore, the resonance cannot take place  in the PPLT configuration considered in Sec.\ref{sec_PPLT}, and from now on we focus on the BBO case only, setting $G_z=0$. \\
 We notice that in principle 
the  r.h.s. of Eq.\eqref{res1} depends on the frequency.  The only exception is when one of the pumps is not tilted, e.g. $\theta_{p1} = 0 $. Then,  by requiring that  shared modes are generated by the other one, i.e. that   $ q_{0x} (\Omega) + q_{0x} (-\Omega) = Q_2$, for $   \frac{\Delta k_p}{\Delta Q_p}  = \frac{\theta_{p2}}{2} =\frac{\theta_{p1} +\theta_{p2}}{2} $  the resonance takes place simultaneously at all the frequencies. However, even when this "magic" configuration is not considered, the bandwidth of modes that enter into resonance is so huge that can be practically considered infinite. We assume that Eq.\eqref{res1} is satisfied at degeneracy where  $ \Dcoll (0) =0$, i.e. that 
\beq
\frac{\Delta k_p}{\Delta Q_p}=\left(\frac{\Delta k_p}{\Delta Q_p}\right)_{res}  =\pm 
\frac{\theta_{p1} - \theta_{p2}} {2} 
\label{res}
\eeq
Then, at a frequency $\Omega \ne 0$ the relative correction to the resonance condition in Eq.\eqref{res1}  is on the order 
$\frac{ \Dcoll (\Omega) } {k_p} \approx \frac{1}{k_p} k''_s \Omega^2  =\frac{\Omega^2}{\Omega_B^2} $, 
where $\Omega_B = \sqrt{ \frac{k_p}{k''_s} } \approx 2 \times  10^{16} \, \mathrm{s}^{-1}$. Thus, for any practical purpose, condition \eqref{res} can be taken as {\em  the} resonance condition. 
\par 
A further insight into the problem is gained by approximating the incremental ratio in Eq. \eqref{dkdqA} with its Taylor expansion. It turns out that the lowest order  approximation  is not precise enough,  therefore we choose to expand each $k_{pj}$   around the middle point $\bar Q_p = {Q_{1} + Q_{2} \over 2}$ as 
$k_{p2,p1} =  k_p (\bar Q_p)  \pm \left. \frac{d k_p}{d Q } \right|_{\bar Q_p }  \frac{\Delta Q_p  }{2} + {1 \over 8} 
\left. \frac{d^2 k_p}{d Q^2 } \right|_{\bar Q_p}  \Delta Q_p ^2  + O(\Delta Q_p^3)
$. In this way,  $\frac{\Delta k_p}{\Delta Q_p} 
=  \left. \frac{d k_p}{d Q } \right|_{\bar Q_p }  + O (\Delta Q_p^2)$. Therefore,  up to first order in $\Delta Q_p$ one has
\beq
\frac{\Delta k_p}{\Delta Q_p} %=\frac{k_{p2} -k_{p1}}{Q_2 -Q_1} 
 \simeq  \left. \frac{d k_p}{d Q } \right|_{\bar Q_p } 
=\left. \frac{1}{k_p}  \frac{d k_p}{d \theta_p } \right|_{\bar \theta_p }  
= \left. \frac{1}{k_p}  \frac{d k_p}{d \gamma }  \,   \frac{d \gamma}{d \theta_p }  \right|_{\bar \theta_p }  
\label{dkdq1}
\eeq
where 
 $\bar \theta_p = {\theta_{p1} + \theta_{p2} \over 2}$, and we remind that $\gamma $ is the angle formed by the pump propagation direction with the optical axis. 
 In this expression we recognize  that the quantity 
$  \frac{1}{k_p}  \frac{d k_p}{d \gamma } = - \rho_\gamma$  is the { \em walk-off angle } formed  by the wave-vector of the extraordinary wave and its Poynting vector, representing the direction of the energy flux. \cite{BornWolf1999} It depends  on  the angle $\gamma$, but we make a small error in taking it at the cut angle $\gamma_0$,  $ \rho_\gamma \to \rho_{0}  \simeq 0.0744    \text{ radians} = 4.26^\circ$.
Thus, with a precision up to first order in the small quantities the following expression holds: 
\beq
\frac{\Delta k_p}{\Delta Q_p}   =  -\rho_{\gamma}  \left. \frac{d \gamma}{d \theta_p }  \right|_{\bar \theta_p }  
%\left( \sin \beta - {\theta_{p1} + \theta_{p2} \over 2} \frac{1}{ \tg \gamma_0}\right)
\eeq
On the other side, the functional dependence of the angle $\gamma$  on the tilt angle $\theta_p$  is  provided by Eq.\eqref{gamma}. By differentiating this expression with respect to $\theta_p$, one gets
\beq
\begin{aligned}
\frac{d \gamma}{d \theta_p }  &= - \sin \beta \cos \theta_p  \frac{ \sin \gamma_0 }{\sin \gamma } + \sin \theta_p \frac{ \cos \gamma_0 }{\sin \gamma }   \\
%&\simeq - \sin \beta    + \frac{ \theta_p  }{\tg \gamma_0 }  & \text{for } \theta_p \ll  1
& \to \left\{ \begin{array}{lc}    \mp 1  & \text{for } \beta = \pm  {\pi \over 2} \\
- \sin \beta    + \sin \theta_p \frac{1  }{\tg \gamma_0 } & \text{for } |\beta |\ll  {\pi \over 2} \end{array} \right. 
\end{aligned}
\label{dgammadtheta}
\eeq
The resonance condition of  Eq. \eqref{res} can then be reformulated in terms of the tilt angles of the two pumps as 
\begin{align}
 \pm  \frac{\theta_{p1} - \theta_{p2}} {2}  %&= -\rho_{0}  \left. \frac{d \gamma}{d \theta_p }  \right|_{\bar \theta_p }   \\
&=\left.  \rho_\gamma \left( \sin \beta \cos \theta_p  \frac{ \sin \gamma_0 }{\sin \gamma } - \sin \theta_p \frac{ \cos \gamma_0 }{\sin \gamma }  \right)\right|_{\theta_p = \bar \theta_p} \\
& \simeq \left\{  \begin{array} {lc} + \rho_{\bar \gamma}      & \quad \beta =+ \frac{\pi}{2}  \\
-\rho_{\bar \gamma}      & \quad \beta = -\frac{\pi}{2}  \\
\rho_{0} \left( \sin \beta - {\theta_{p1} + \theta_{p2} \over 2} \frac{1}{ \tg \gamma_0}\right) & \quad | \beta|  \ll \frac{\pi}{2}  
\end{array} \right. 
\qquad 
\end{align}
This condition can be understood as a requirement on the pump tilt angles, for a fixed angle of rotation $\beta$  of the crystal, or viceversa, for given pump tilts $\theta_{p1}, \theta_{p2}$ as a receipt for  the angle of rotation of the crystal at which resonance takes place. 
\beq
\sin(\beta^{\mathrm res})=\pm   \frac{\theta_{p1} - \theta_{p2}}{2  \rho_0 } +  \frac{\theta_{p2} +  \theta_{p1}}{2 \tg \gamma_0 }
\label{betares}
\eeq
%%%%%%%%%%%%%%%%%%%%%%
%\bibliography{biblio_BulkDual2020}
%\bibliographystyle{apsrev4-1}

\end{document}